\documentclass[12pt]{article}
\usepackage{epsfig,cite}
\usepackage {amsmath}
\usepackage{amssymb}
\include{epsf}
\newcount\timecount
\newcount\hours \newcount\minutes  \newcount\temp \newcount\pmhours
\hours = \time
\divide\hours by 60
\temp = \hours
\multiply\temp by 60
\minutes = \time
\advance\minutes by -\temp
\def\hour{\the\hours}
\def\minute{\ifnum\minutes<10 0\the\minutes
            \else\the\minutes\fi}
\def\clock{
\ifnum\hours=0 12:\minute\ AM
\else\ifnum\hours<12 \hour:\minute\ AM
      \else\ifnum\hours=12 12:\minute\ PM
            \else\ifnum\hours>12
                 \pmhours=\hours
                 \advance\pmhours by -12
                 \the\pmhours:\minute\ PM
                 \fi
            \fi
      \fi
\fi
}

\def\monthname{\relax\ifcase\month 0/\or January\or February\or
   March\or April\or May\or June\or July\or August\or September\or
   October\or November\or December\else\number\month/\fi}

\def\bold#1{\setbox0=\hbox{$#1$}%
     \kern-.025em\copy0\kern-\wd0
     \kern.05em\copy0\kern-\wd0
     \kern-.025em\raise.0433em\box0 }

\def\beq{\begin{equation}}
\def\eeq{\end{equation}}

\def\ga{\mathrel{\raise.3ex\hbox{$>$\kern-.75em\lower1ex\hbox{$\sim$}}}}
\def\la{\mathrel{\raise.3ex\hbox{$<$\kern-.75em\lower1ex\hbox{$\sim$}}}}
\def\gev{{\rm \, Ge\kern-0.125em V}}
\def\tev{{\rm \, Te\kern-0.125em V}}
\def\gyr{{\rm \, G\kern-0.125em yr}}

\def\tbt{\tan \beta}

\def\gappeq{\mathrel{\rlap {\raise.5ex\hbox{$>$}}
{\lower.5ex\hbox{$\sim$}}}}
\def\lappeq{\mathrel{\rlap{\raise.5ex\hbox{$<$}}
{\lower.5ex\hbox{$\sim$}}}}
\def\Toprel#1\over#2{\mathrel{\mathop{#2}\limits^{#1}}}

\def\stau{\widetilde \tau}

\def\m12{m_{1\!/2}}

\def\PL{{Phys.~Lett.} }

\def\stau{\tilde{\tau}}

\def\bea{\begin{eqnarray}}
\def\eea{\end{eqnarray}}
\newcommand{\goto}{\rightarrow}

\begin{document}
\begin{titlepage}
\pagestyle{empty}
\baselineskip=21pt
\rightline{CERN-PH-TH/2008-004}
\rightline{UMN--TH--2630/08}
\rightline{FTPI--MINN--08/01}
\vskip 0.2in
\begin{center}
{\large{\bf Sparticle Discovery Potentials in the CMSSM and GUT-less 
Supersymmetry-Breaking Scenarios}}
\end{center}
\begin{center}
\vskip 0.2in
{\bf John~Ellis}$^1$, {\bf Keith~A.~Olive}$^{2}$ and
{\bf Pearl Sandick}$^{2}$
\vskip 0.1in

{\it
$^1${TH Division, PH Department, CERN, CH-1211 Geneva 23, Switzerland}\\
$^2${William I. Fine Theoretical Physics Institute, \\
University of Minnesota, Minneapolis, MN 55455, USA}\\
}

\vskip 0.2in
{\bf Abstract}
\end{center}
\baselineskip=18pt \noindent

We consider the potentials of the LHC and a linear $e^+e^-$ collider (LC) for discovering
supersymmetric particles in variants of the MSSM with soft supersymmetry-breaking mass
parameters constrained to be universal at the GUT scale (CMSSM) or at some lower scale $M_{in}$
(GUT-less models), as may occur in some scenarios with mirage unification. Whereas the LHC should 
be able to discover squarks and/or gluinos along all the CMSSM coannihilation strip where the relic
neutralino LSP density lies within the range favoured for cold dark matter, many GUT-less
models could escape LHC detection. In particular, if $M_{in} < 10^{11}$~GeV, the LHC would not
detect sparticles if the relic density lies within the favoured range. For any given discovery of
supersymmetry at the LHC, in such GUT-less models the
lightest neutralino mass and hence the threshold for sparticle pair production at a LC
{\it increases} as $M_{in}$ {\it decreases}, and the CMSSM offers the best prospects for
measuring sparticles at a LC. For example, if the LHC discovers sparticles with 1~fb$^{-1}$ of 
data, within the CMSSM a centre-of-mass energy of 600~GeV would suffice for  a LC to to produce 
pairs of neutralinos, if they provide the cold dark matter, whereas over 1 TeV might be required in 
a general GUT-less model. These required energies increase to 800~GeV in the CMSSM
and 1.4~TeV in GUT-less models if the LHC requires 10~fb$^{-1}$ to discover supersymmetry.

\vfill
\leftline{CERN-PH-TH/2008-004}
\leftline{January 2008}
\end{titlepage}

\section{Introduction}

Many studies have showcased the great potential of the LHC for
producing and discovering supersymmetric particles \cite{lhc,cmstdr,LHC}, and the ability
of experiments at a linear $e^+ e^-$ collider (LC) to measure sparticle properties in detail, if
their pair-production thresholds lie within its kinematic reach \cite{ilc}. Most
of these studies have assumed that
$R$ parity is conserved, in which case the lightest supersymmetric
particle (LSP) may provide the cold dark matter postulated by
astrophysicists and cosmologists \cite{EHNOS}. Further, most studies have been within the
framework of the minimal supersymmetric extension of the Standard Model (MSSM) \cite{mssm},
and assumed that the LSP is the lightest neutralino $\chi$.
We also adopt this framework in this paper.
In this case, the classic signature of sparticle pair production is missing energy
carried away by the dark matter particles $\chi$. Studies have
indicated that experiments at the LHC should be able to detect gluinos and
squarks weighing up to $\sim 2.5$~TeV \cite{Baer}, whereas any sparticles weighing
less than the beam energy should be detectable at a LC.

One specific supersymmetric version of this framework that has
commonly been examined is the Constrained MSSM 
(CMSSM)~\cite{funnel,cmssm,efgosi,eoss,cmssmwmap}, in which the
soft supersymmetry-breaking mass parameters are assumed to be universal
at some high scale, generally taken to be the supersymmetric GUT scale, 
$M_{GUT} \sim 10^{16}$ GeV.  Within the CMSSM,
renormalization group equations (RGEs) can be used to
calculate the weak-scale observables in terms of four continuous and one
discrete parameter; the scalar mass, $m_0$, the gaugino mass,
$m_{1/2}$, and the trilinear soft breaking parameter, $A_0$ (each
specified at the universality scale), as well as the ratio of the Higgs vevs, $\tbt$, and the
sign of the Higgs mixing parameter, $\mu$. The reaches of colliders such as the LHC
or a LC are then often expressed in the $(m_{1/2}, m_0)$ plane for
representative values of $A_0, \tbt$ and the sign of $\mu$.

However, the mechanism of supersymmetry breaking is not known,
and alternative scenarios should also be considered.
Rather than postulate that the soft supersymmetry-breaking
parameters are universal at some GUT scale, one might consider theories
in which this universality assumption for the the soft supersymmetry-breaking 
parameters is relaxed. One possibility, motivated to some extent by
supersymmetric GUT scenarios and the absence of flavour-changing
interactions due to sparticle exchanges, would be to relax (for example)
the universality assumption for the soft supersymmetry-breaking contributions
to the Higgs scalar masses at the GUT scale (the NUHM) \cite{nonu,nuhm}, and more radical
abandonments of universality could also be considered.

We consider here a different generalization of the CMSSM, in which
universality of the soft supersymmetry-breaking mass
parameters is maintained, but is imposed at some 
lower input scale $M_{in} < M_{GUT}$ \cite{eos1,eos2}.
Such GUT-less (or sub-GUT) scenarios may arise in models where the dynamics
that breaks or communicates supersymmetry breaking to the observable sector
has an intrinsic scale below $M_{GUT}$, and switches off at higher scales, much
as the effective dynamical quark mass in QCD switches off at scales $> \Lambda_{QCD}$.
Mirage unification scenarios~\cite{mixed} offer one class of examples in which the 
low-energy evolution of the gaugino masses is as if they unified at some scale $< M_{GUT}$.
In principle, one could consider scenarios in which universality is imposed on the 
different MSSM soft 
supersymmetry breaking parameters $m_{1/2}, m_0$ and $A_0$ at different input
scales $M_{in}$. However, here we follow~\cite{eos1,eos2} in studying the simplest
class of GUT-less scenarios with identical $M_{in}$ for all the soft supersymmetry-breaking
parameters. 

As one would expect, the reduction in the universality scale has important
consequences for the low-energy sparticle mass spectrum. In particular,
the hierarchy of gaugino masses familiar in the GUT-scale CMSSM is
reduced with, for example, a substantial reduction in the ratio of gluino
and bino masses. Likewise, squark and slepton masses also approach each
other as $M_{in}$ is reduced. These effects have important consequences
for the $(m_{1/2}, m_0)$ planes in GUT-less scenarios: for example, the boundaries
imposed by the absence of a charged ${\tilde \tau_1}$ LSP and the generation of an
electroweak symmetry breaking vacuum approach each other as $M_{in}$
decreases.

A corollary of the `squeezing' of the sparticle mass spectrum is the
observation made in \cite{eos1} and \cite{eos2} that, as the universality
scale $M_{in}$ is decreased from the GUT scale, there are dramatic
changes in the cosmological constraint imposed on the parameter space by the
relic density of neutralinos inferred from WMAP and other observations~\cite{WMAP}. In
general, as $M_{in}$ decreases, the regions where the relic neutralino LSP density
falls within the range preferred by WMAP and other measurements~\cite{WMAP} tend to move to larger
$m_{1/2}$ and $m_0$. This implies that, whereas in the GUT-scale CMSSM
the relic neutralino is {\it overdense} in most of the region with $m_{1/2}, m_0 < 1$~TeV,
as $M_{in}$ decreases to $\sim 10^{11}$~GeV most of this region becomes
{\it underdense}.

In this paper, we consider the implications of these observations for the prospects
for sparticle detection at the LHC and a LC. ATLAS and CMS have
estimated their reaches in inclusive supersymmetry searches for multiple jets and 
missing transverse energy, as functions of the accumulated and analyzed LHC
luminosity, which may be expressed as reaches for gluino and
squark masses~\cite{cmstdr}. These may in turn be converted into the reaches in the
$(m_{1/2}, m_0)$ planes for different values of $M_{in}$. The masses of
weakly-interacting sparticles such as sleptons, charginos and neutralinos are
determined across these $(m_{1/2}, m_0)$ planes, and hence the ATLAS/CMS
reaches may be converted into the corresponding sparticle pair-production
thresholds at a generic LC. These converted reaches may be
interpreted in at least two ways. If the LHC {\it does discover} supersymmetry,
then one may estimate, within the CMSSM or any given GUT-less model, the {\it maximum}
centre-of-mass energy that would suffice for a LC to make detailed follow-up
measurements of at least some sparticles. Conversely, if the LHC
{\it does not discover} supersymmetry within a given physics reach,
one can, within the CMSSM or any given GUT-less model, estimate the {\it minimum}
centre-of-mass energy below which a LC would not provide access to any
sparticles. In general, because of the `squeezing' of the sparticle mass spectrum as $M_{in}$
{\it decreases}, for any given LHC physics reach
the required LC centre-of-mass energy {\it increases}
correspondingly.

This argument can be carried through whether one disregards the cosmological
density of dark matter entirely, or regards it solely as an upper limit on the relic
LSP density, or interprets it as a narrow preferred band. In the third case, the
prospects for sparticle detection at the LHC recede with the preferred 
dark matter regions in
the $(m_{1/2}, m_0)$ planes as $M_{in}$ decreases. Within the specific
preferred dark-matter regions, the relation between the LHC and LC reaches can be made more 
precise. For example, if the LHC discovers sparticles with 1~fb$^{-1}$ of 
data, within the CMSSM a centre-of-mass energy of 600~GeV would suffice for  a LC to to produce 
pairs of neutralinos, if they provide the cold dark matter, whereas over 1 TeV might be required in 
a GUT-less model with $M_{in} > 10^{11.5}$~GeV. These required energies increase to 800~GeV 
in the CMSSM and 1.4~TeV in GUT-less models with $M_{in} > 10^{11.5}$~GeV if the LHC requires 
10~fb$^{-1}$ to discover supersymmetry.

\section{Sparticle Masses in GUT-less Models}

\begin{figure}[ht!]
\begin{center}
\mbox{\epsfig{file=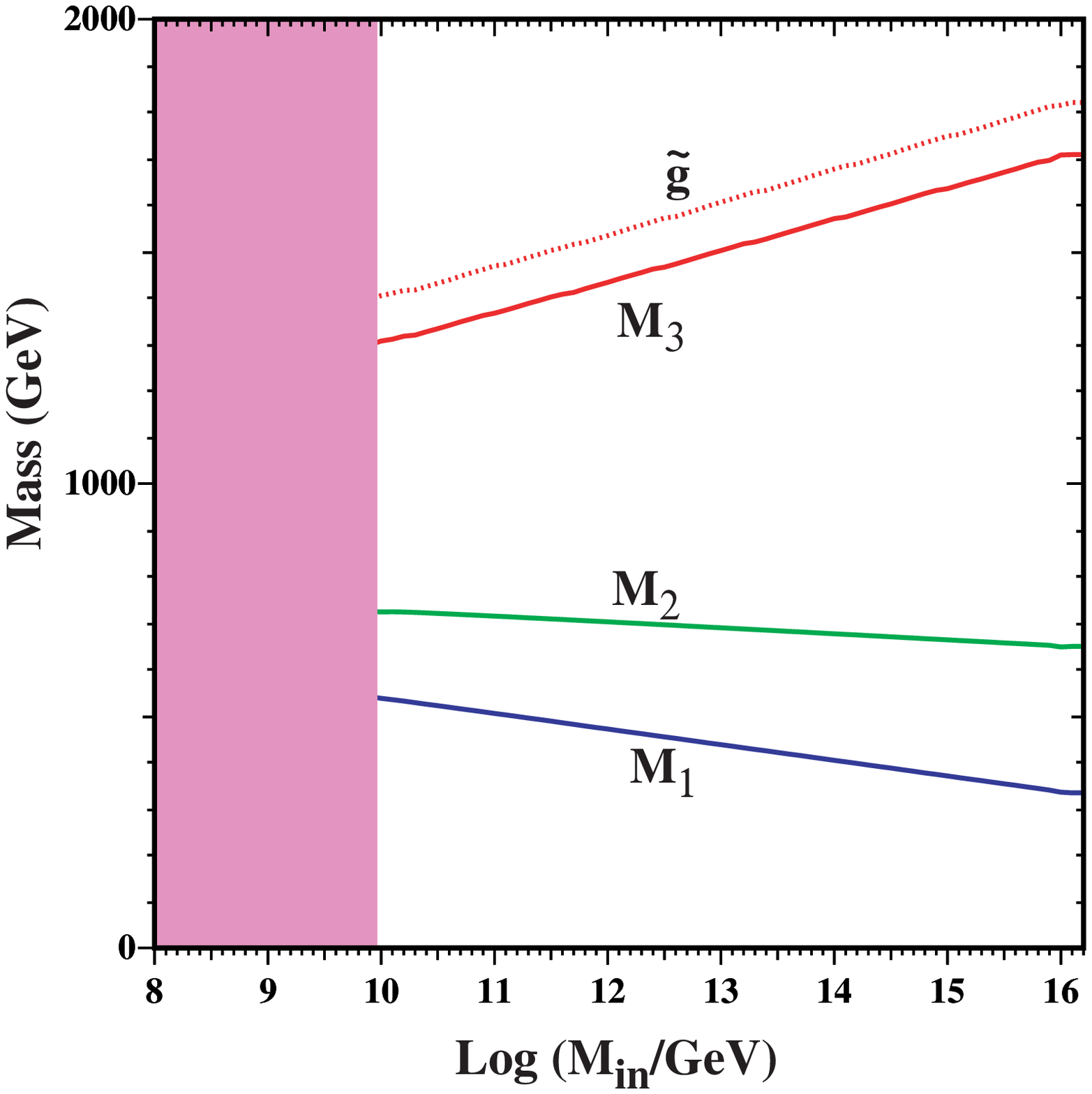,height=6.8cm}}
\mbox{\epsfig{file=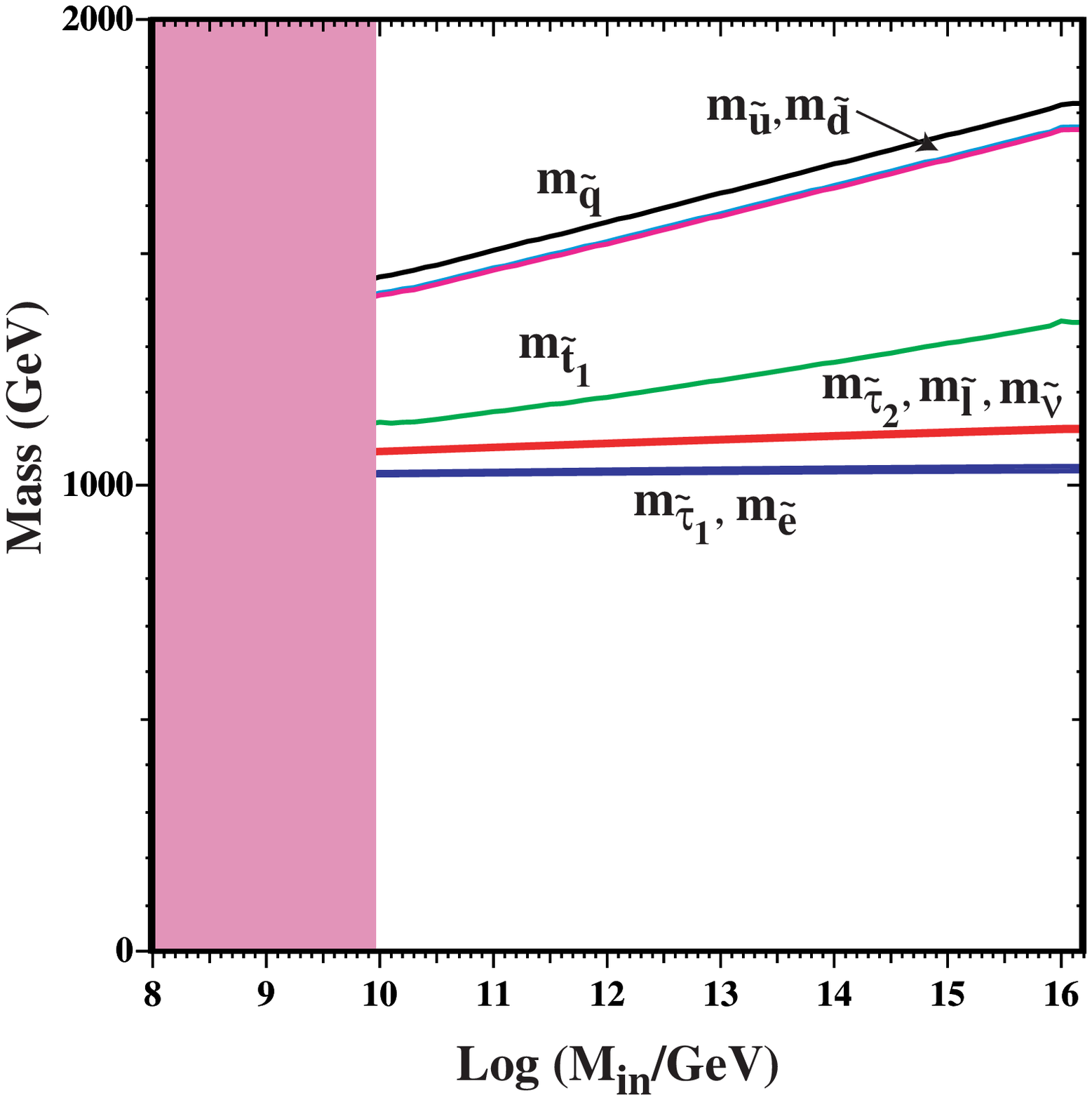,height=6.8cm}}
\end{center}
\caption{\it Panel (a) shows the low-energy effective gaugino masses as functions of
$M_{in}$ for the point $(m_{1/2},m_0) = (800,1000)$ GeV, with
$\tan{\beta} = 10$, $A_0 = 0$, and $\mu > 0$.
Panel (b) shows the corresponding dependence on $M_{in}$ of the squark and
slepton masses as indicated for the same value of $(m_{1/2},m_0)$.
\label{fig:masses}}
\end{figure}

Before discussing in depth the physics reaches of different colliders, we first discuss
the behaviours of some relevant sparticle masses in GUT-less scenarios,
starting with the gauginos. Since the leading one-loop renormalization-group
evolutions of the gaugino masses $M_a(Q): a = 1,2, 3$
are identical with those of the gauge coupling strengths $\alpha_a(Q)$,
\begin{equation}
\label{eq:gauginos}
M_a(Q) = \frac{\alpha_a(Q)}{\alpha_a(M_{in})}m_{1/2}.
\end{equation}
At the one-loop level, the running gaugino masses therefore
track the behaviours of the gauge couplings, and
$\alpha_a(Q)/\alpha_a(M_{in}) \goto 1$ as $M_{in} \goto Q$. Since the SU(3) gauge coupling is asymptotically free
whereas the SU(2) and U(1) couplings increase with the
renormalization scale, it is clear that the running gluino
mass at the electroweak scale {\it decreases} towards $m_{1/2}$ as $M_{in}$ is decreased,
whereas the running wino and bino masses {\it increase} towards $m_{1/2}$
as one approaches $M_{in}$.
At the two-loop level, the renormalizations of the
gaugino masses and the gauge couplings are different, but the
one-loop effect (Eq.~\ref{eq:gauginos}) is clearly dominant, as seen in
panel (a) of Fig.~\ref{fig:masses} for the representative case
$m_{1/2} = 800$~GeV~\footnote{All of the results presented here include 
two-loop effects in the RGEs.}. As $M_{in}$ decreases,
$M_3$ decreases and $M_{1,2}$ increase towards the input value
$m_{1/2} = 800$~GeV. 

The physical gaugino masses differ from the running masses
by threshold corrections at the electroweak scale, of which the most important is that
for the gluino mass. At the one-loop level, this correction takes the form
\begin{equation}
m_{\widetilde g} \; = \; M_3(Q) \, - \, {\rm Re} \Sigma_{\widetilde g} \, m_{\widetilde g}^2,
\label{runphys}
\end{equation}
where $\Sigma_{\widetilde g} m_{\widetilde g}^2$ incorporates the effects
due to gluon-gluino and quark-squark loops \cite{pbmz}. These effects often amount
to $\sim 10$~\%, as also shown in panel (a) of Fig.~\ref{fig:masses} for the 
representative case $m_{1/2} = 800$~GeV, where the one-loop threshold
corrections are calculated assuming $m_0 = 1000$~GeV, $A_0 = 0$
and $\tbt = 10$ at $M_{in}$. These electroweak threshold corrections are included in our
subsequent analysis of the physics reaches of the LHC and ILC.

We also include the leading renormalizations of the sfermion masses.
At the one-loop RGE level, the running squark masses may be written as
\begin{equation}
m_{\widetilde{q}}^2(Q) = m_0^2(M_{in}) + C_{\widetilde{q}}(Q,M_{in}) \, m_{1/2}^2,
\label{eq:squark}
\end{equation}
where $C_{\widetilde{q}}$ is a coefficient that decreases with $M_{in}$
for any fixed $Q < M_{in}$, and vanishes as $M_{in} \rightarrow Q$.
Thus, the squark and slepton masses also tend to approach each other and
$m_0$ as $M_{in}$ decreases, {\it modulo} Yukawa corrections and one-loop electroweak 
threshold effects, which we include for stop and sbottom squarks. The dependences of
some squark and slepton masses on $M_{in}$ is shown in panel (b) of Fig.~\ref{fig:masses}.
Whereas the masses of the left- (${\tilde q}$)
and right-handed squarks (${\tilde u}, {\tilde d}$) of the first two generations do tend to
unify with those of the sleptons (${\tilde l}, {\tilde \nu}, {\tilde e}$) as $M_{in}$
decreases, there are important
Yukawa corrections for the lighter stop (${\tilde t_1}$) and sbottom (${\tilde b_1}$),
and smaller corrections for the lighter stau (${\tilde \tau_1}$).

\begin{figure}[p]
\begin{center}
\mbox{\epsfig{file=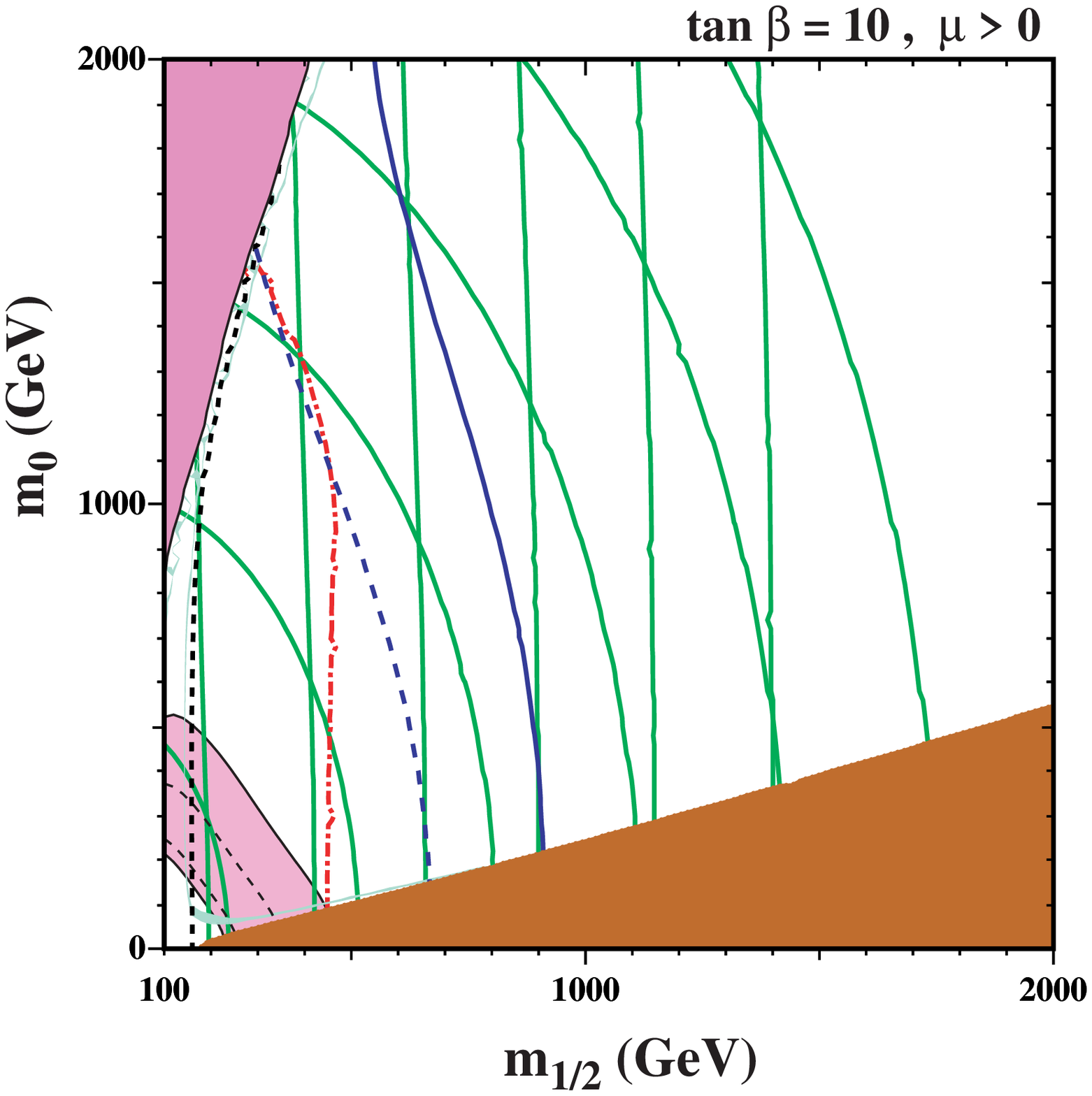,height=6.8cm}}
\mbox{\epsfig{file=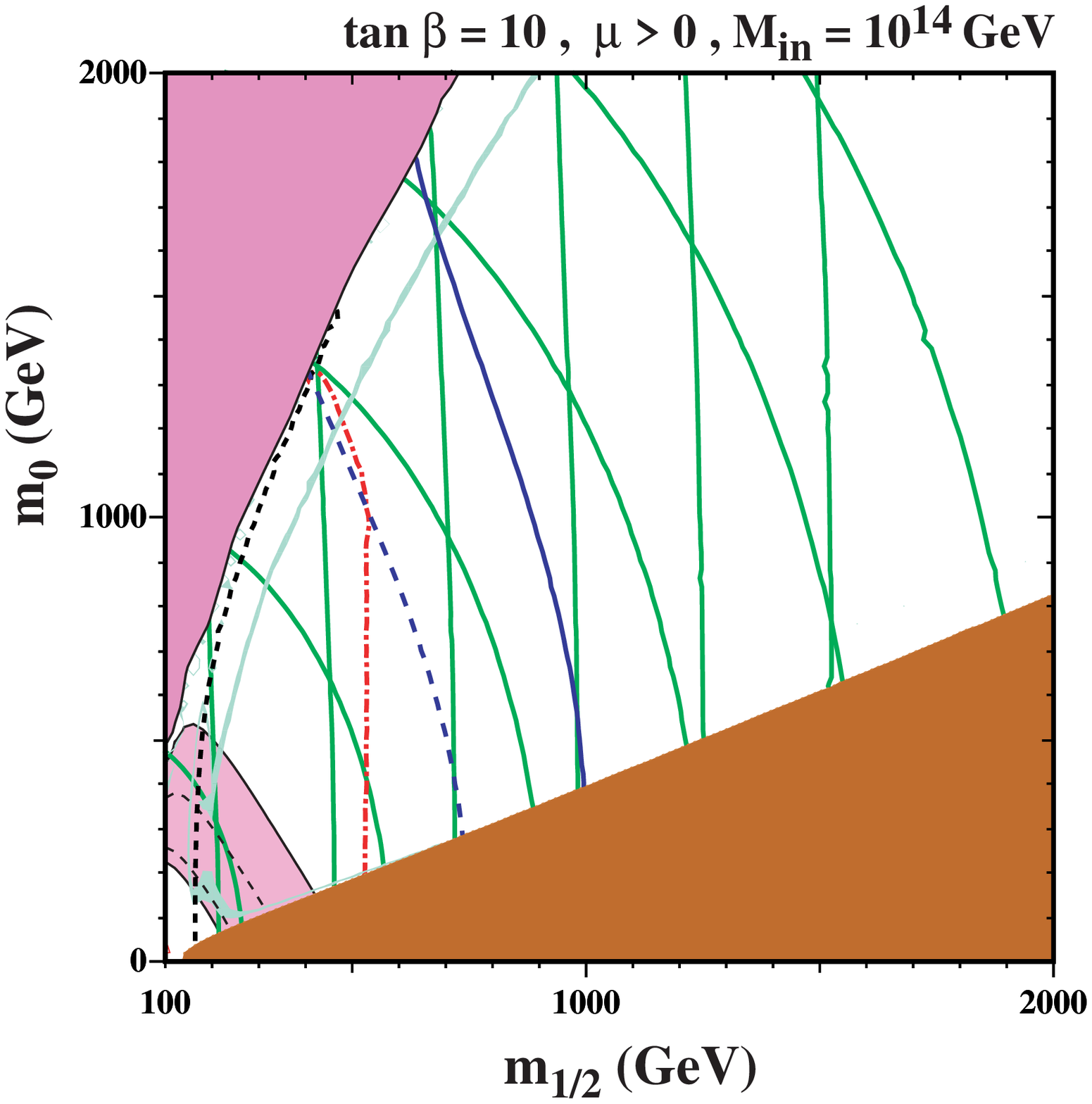,height=7cm}}
\end{center}
\begin{center}
\mbox{\epsfig{file=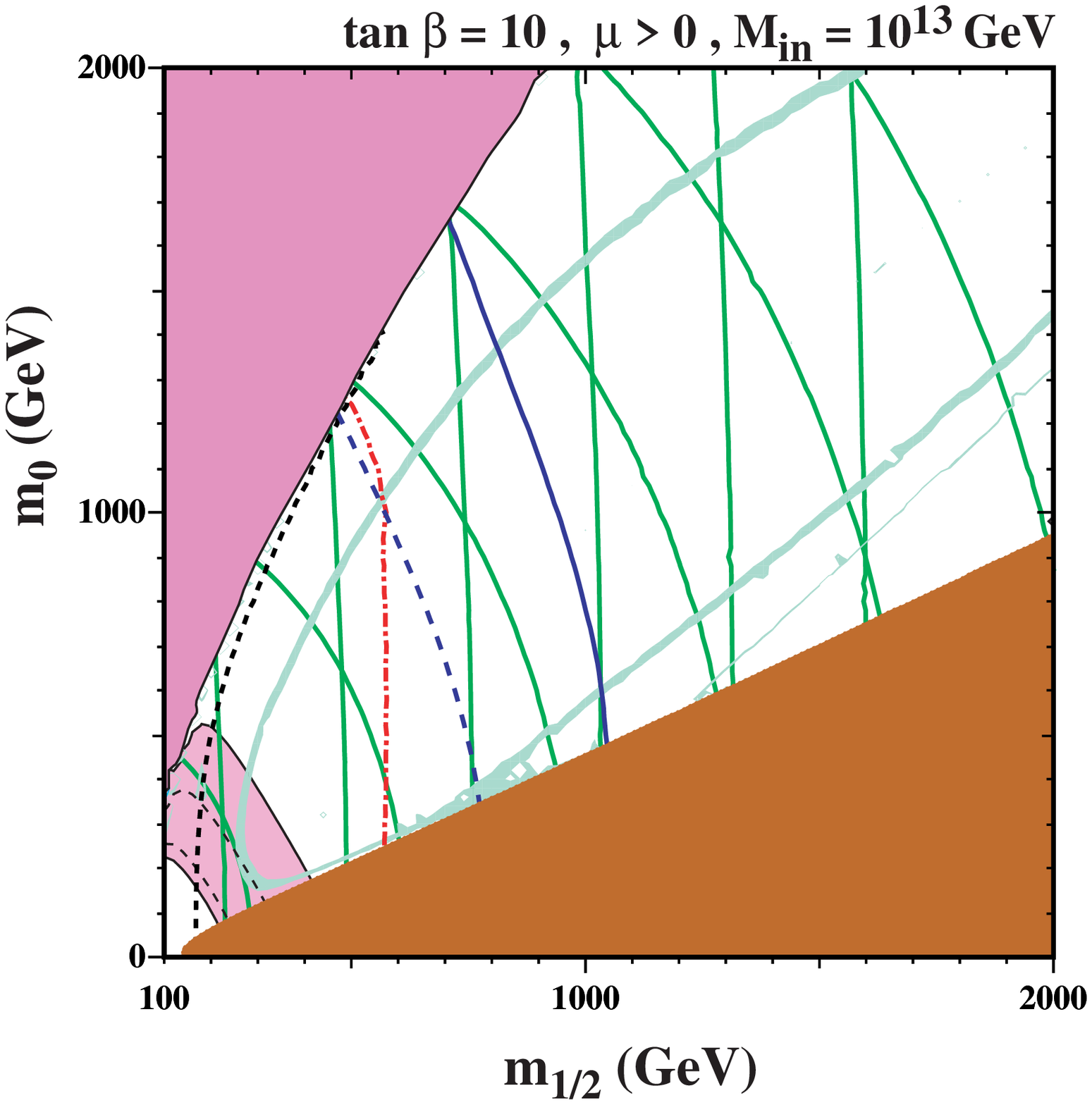,height=7cm}}
\mbox{\epsfig{file=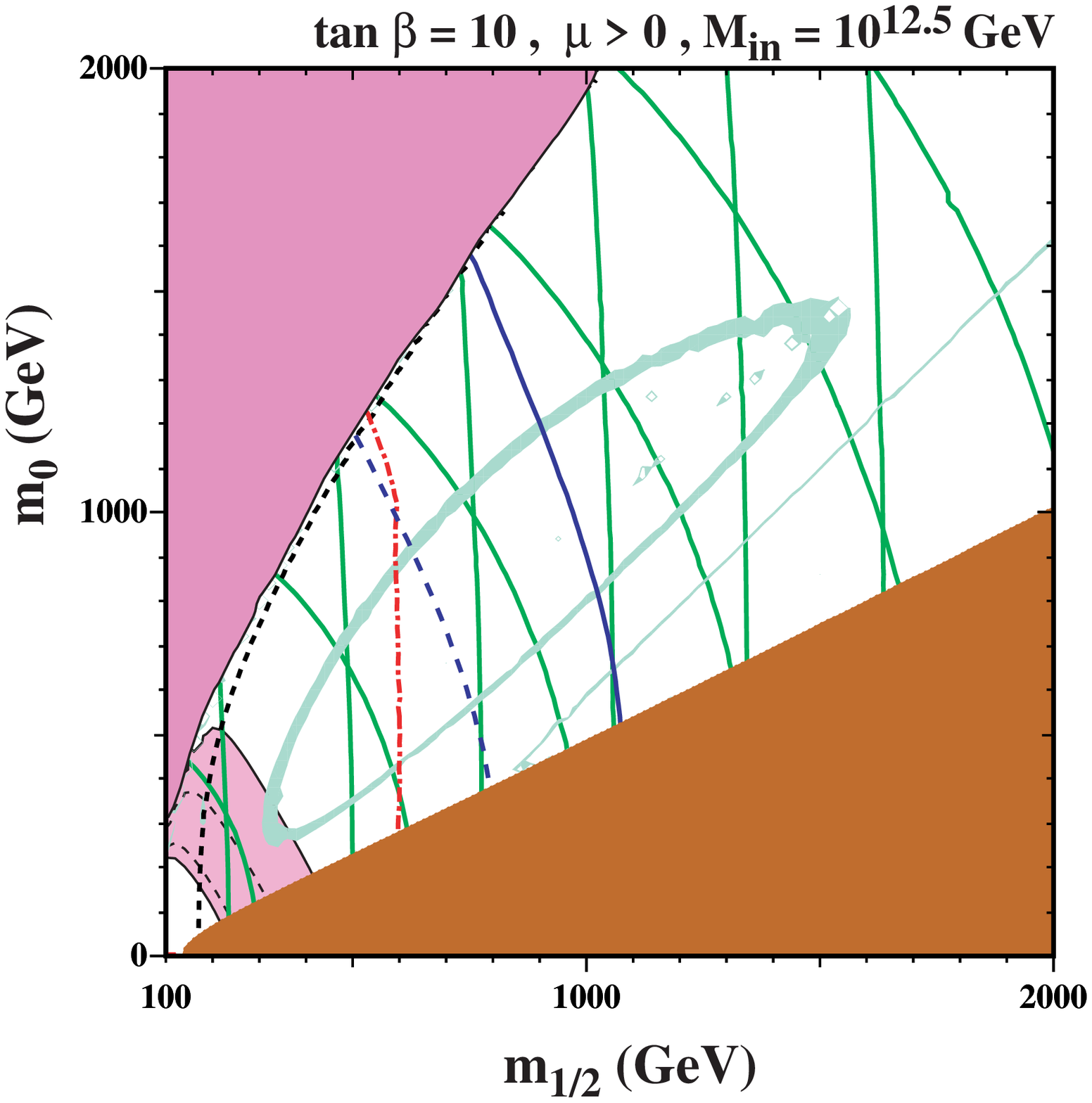,height=7cm}}
\end{center}
\caption{\it Examples of $(m_{1/2}, m_0)$ planes with
$\tan \beta = 10$ and $A_0 = 0$, and (a) $M_{in} = M_{GUT}$, (b) $M_{in} = 10^{14}$~GeV, 
(c) $M_{in} = 10^{13}$~GeV, and (d)
$M_{in} = 10^{12.5}$~GeV. The usual collider
and cosmological constraints are displayed as described in the text.  
In addition, the solid (green) partial ellipses are contours of ${\widetilde{d}_R}$ masses
corresponding to masses of 0.5 - 3~TeV, in 0.5 TeV increments, and the near-vertical (green)
contours are the analogous gluino mass contours. The solid (dashed) dark blue contours correspond 
to the approximate sparticle reach with 10 (1.0)~fb$^{-1}$ of integrated LHC luminosity, as discussed
in the text. 
\label{fig:planes}}
\end{figure}

\begin{figure}[p]
\begin{center}
\mbox{\epsfig{file=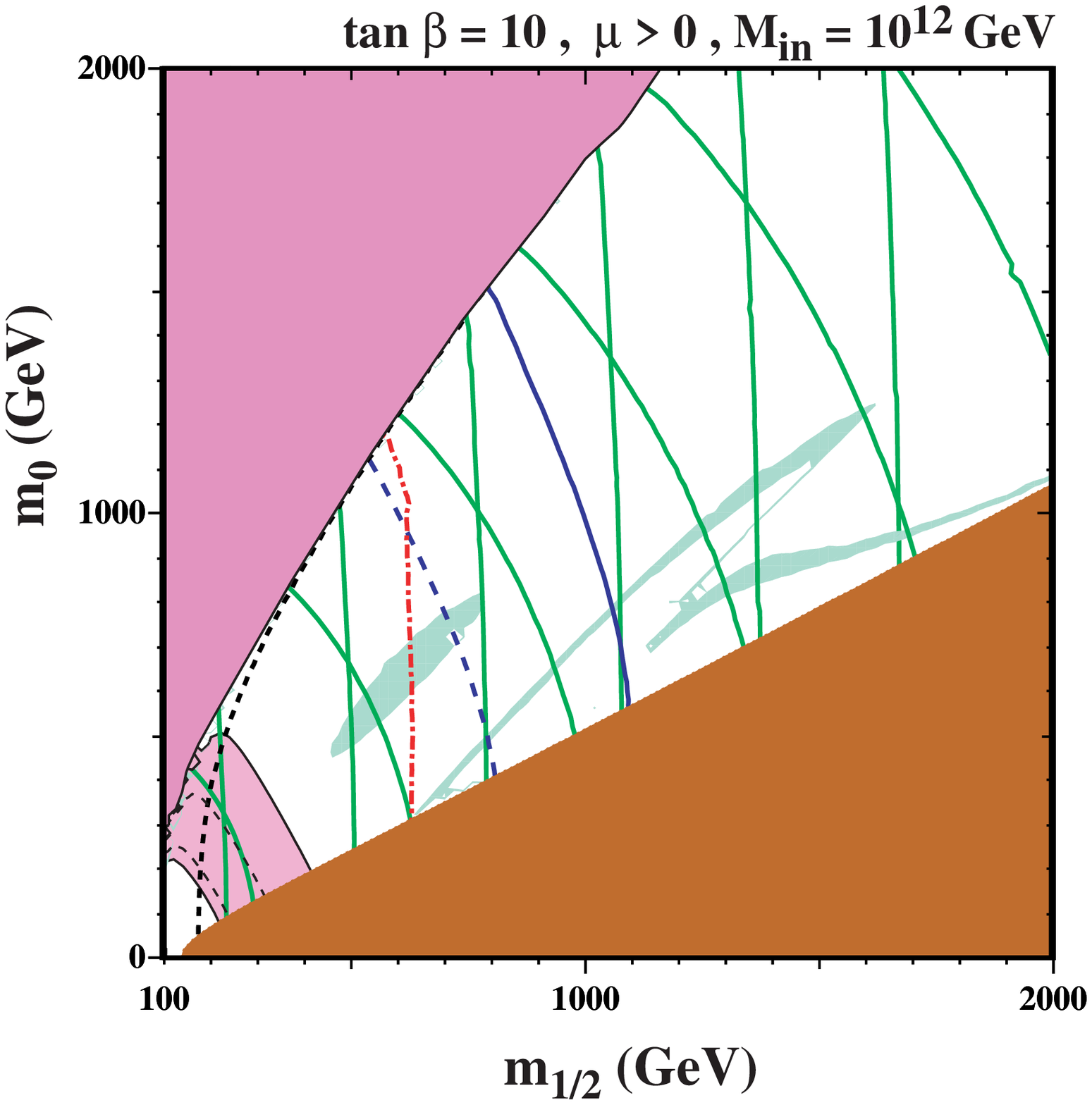,height=6.8cm}}
\mbox{\epsfig{file=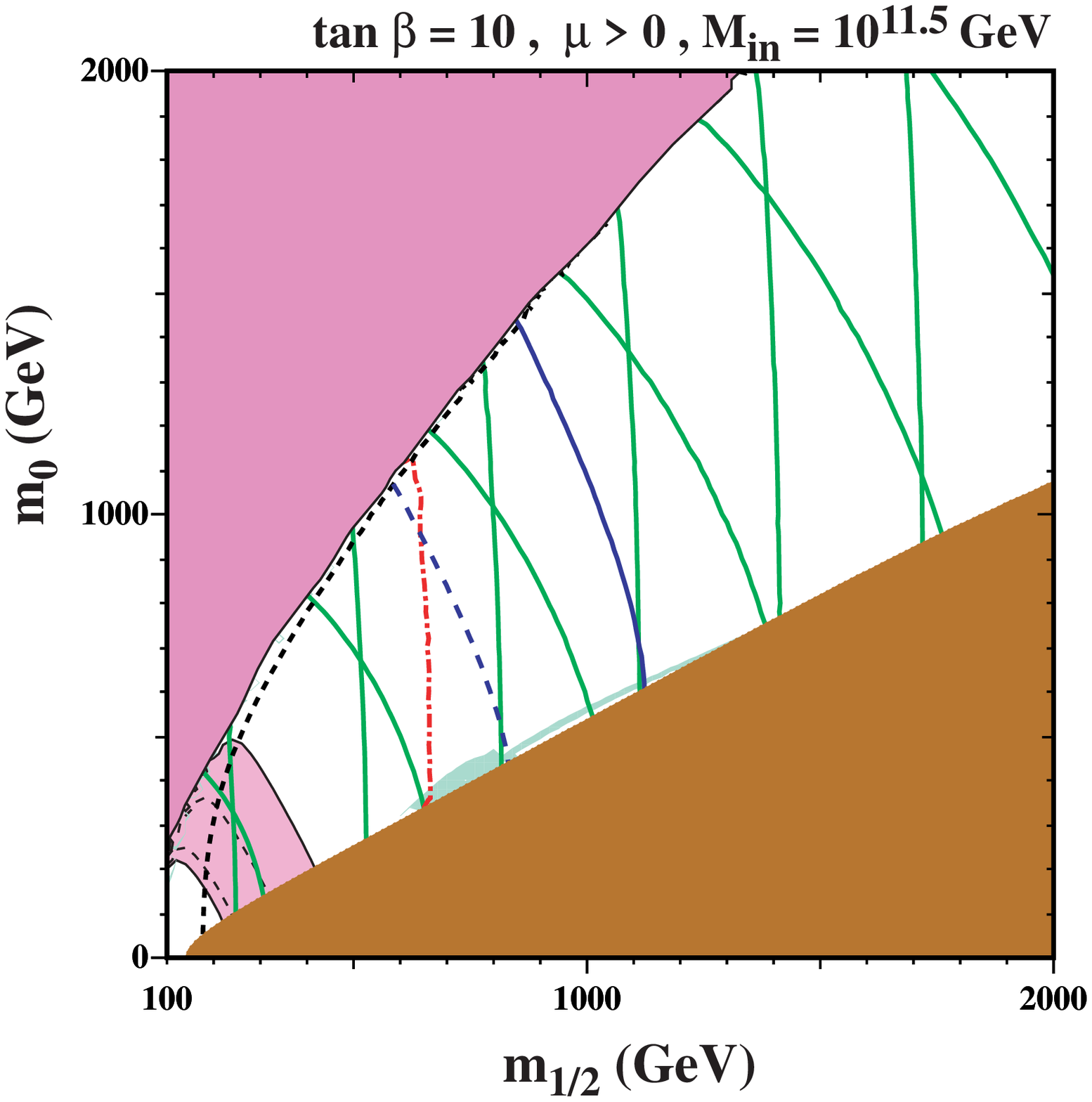,height=7cm}}
\end{center}
\begin{center}
\mbox{\epsfig{file=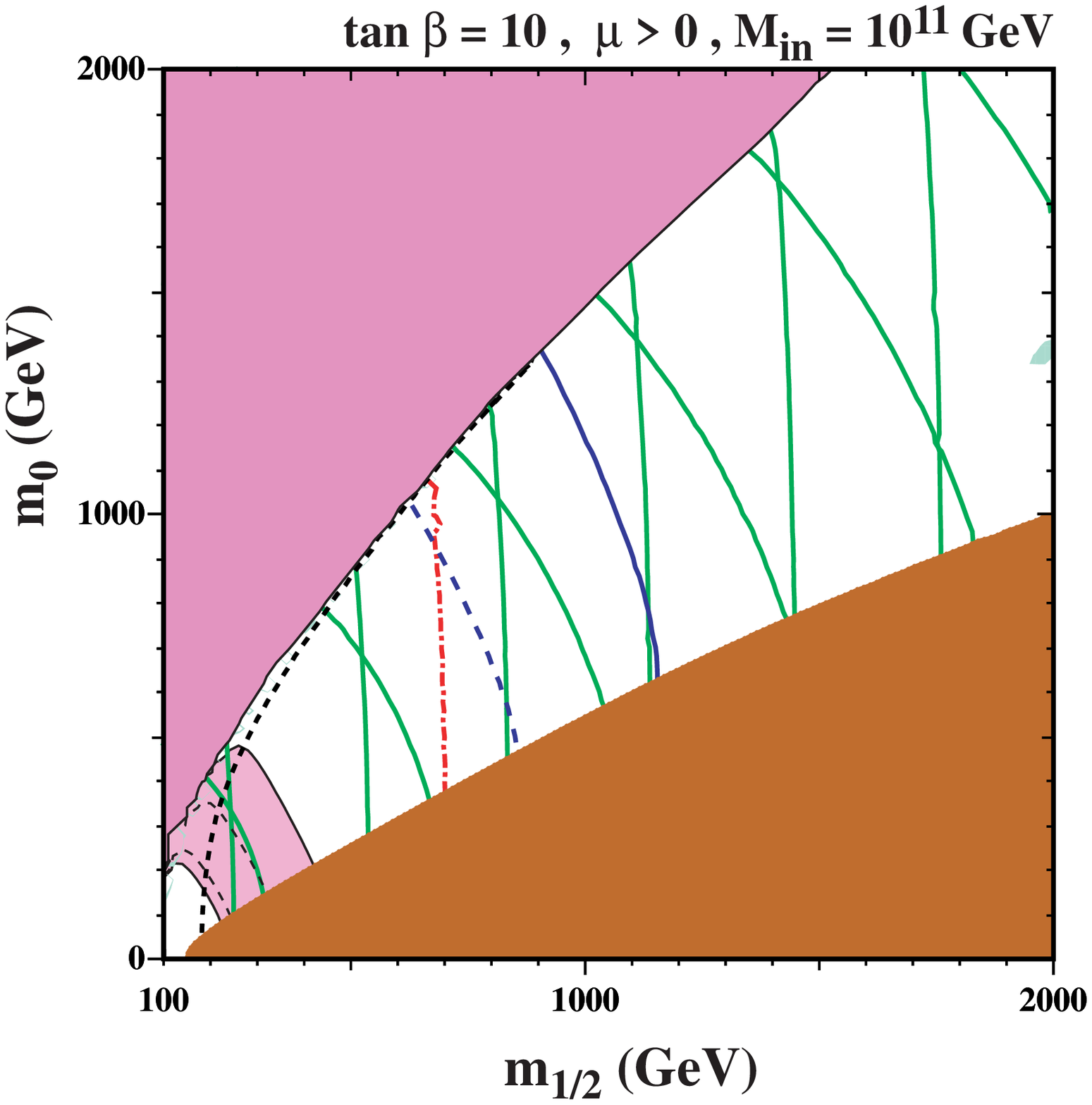,height=7cm}}
\mbox{\epsfig{file=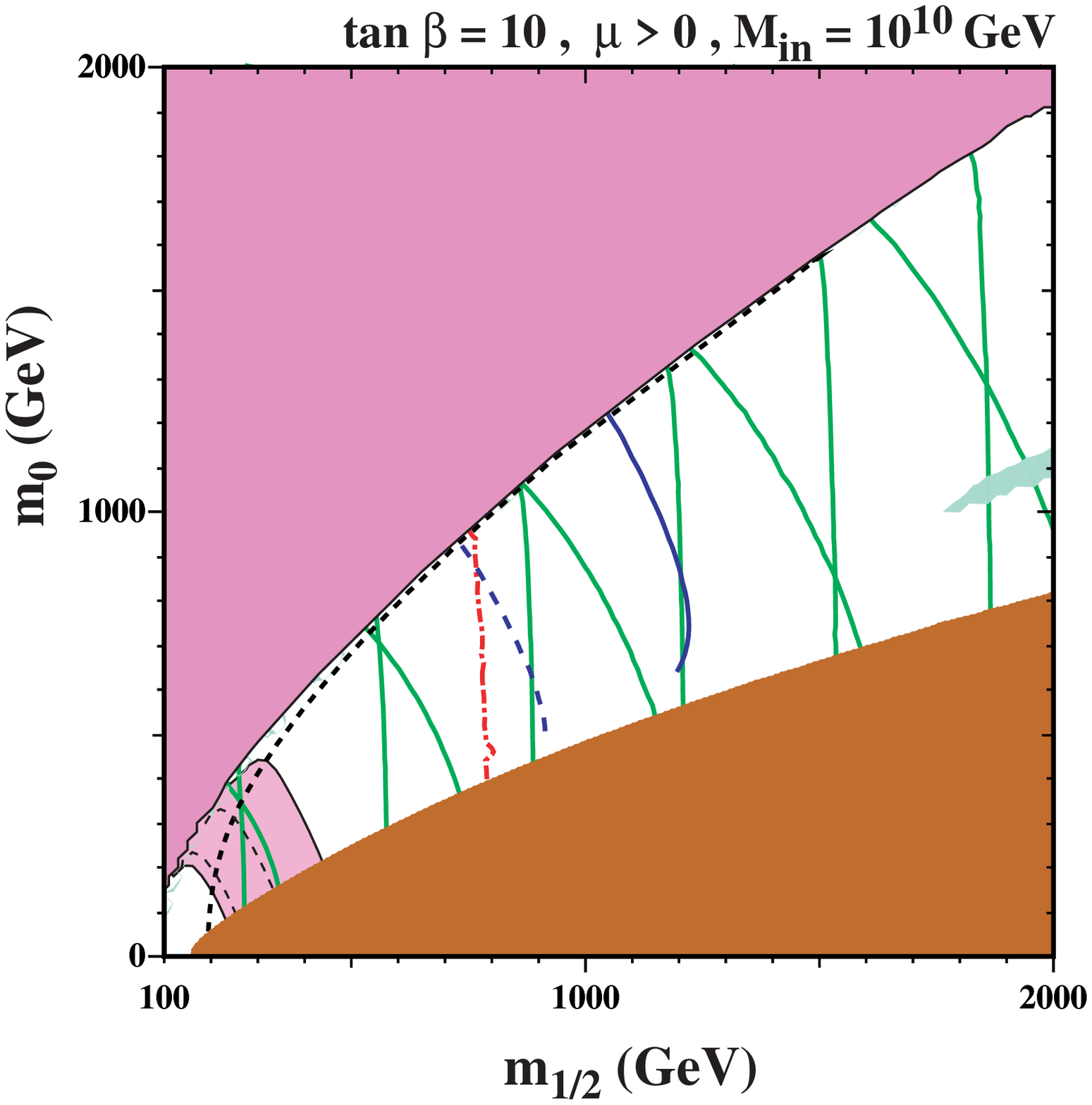,height=7cm}}
\end{center}
\caption{\it Further examples of $(m_{1/2}, m_0)$ planes with
$\tan \beta = 10$ and $A_0 = 0$, and (a) $M_{in} = 10^{12}$, (b) $M_{in} = 10^{11.5}$~GeV, 
(c) $M_{in} = 10^{11}$~GeV, and (d)
$M_{in} = 10^{10}$~GeV. The notations are the same as in Fig.~\protect\ref{fig:planes}.
\label{fig:moreplanes}}
\end{figure}

In preparation for the discussion in the next Section, we display in
Fig.~\ref{fig:planes} the $(m_{1/2},m_0)$ planes for
$\tbt = 10$ and $A_0 = 0$, for various different choices of $M_{in}$:
(a) $M_{GUT}$, (b) $M_{in} = 10^{14}$ GeV, (c) $M_{in} = 10^{13}$ GeV, and (d)
$M_{in} = 10^{12.5}$ GeV, respectively. Further $(m_{1/2},m_0)$ planes for
$\tbt = 10$, $A_0 = 0$ and (a) $M_{in} = 10^{12}$~GeV, (b) $M_{in} = 10^{11.5}$ GeV, 
(c) $M_{in} = 10^{11}$ GeV, and (d) $M_{in} = 10^{10}$ GeV, respectively, are
shown in Fig.~\ref{fig:moreplanes}. Shaded (brown) regions at small $m_0$ and
large $m_{1/2}$ are excluded because the ${\tilde \tau_1}$ is the LSP
whereas shaded (dark pink) regions at large $m_0$ and small $m_{1/2}$ are 
excluded because the electroweak vacuum conditions cannot be met. We note
that these regions approach each other as $M_{in}$ decreases in the successive panels of 
Figs.~\ref{fig:planes} and \ref{fig:moreplanes}. 
Only regions to the right of and below the back dashed lines are
compatible with the LEP constraint on the lightest chargino mass, and only regions to the
right of the red dot-dashed line are compatible with the LEP Higgs mass constraint. The pale
pink shaded bands at small $m_{1/2}$ and $m_0$ are favoured by $g_\mu - 2$
at the one-$\sigma$ level (dashed lines) and two-$\sigma$ level (solid lines) if $e^+ e^-$
data are used to evaluate the Standard Model contribution.

According to (\ref{eq:squark}), squark mass contours may be represented as
approximate quarter-ellipses in the $(m_{1/2},m_0)$ planes, and we show in each
panel as solid (green) lines the contours for
$m_{\widetilde{d}_R}$ = 0.5 - 3~TeV in 0.5~TeV increments. The semimajor axes of 
the quarter-ellipses are approximately equal to $m_{\widetilde{q}}$, and
the semiminor axes are approximately 
equal to $m_{\widetilde{q}}^2/C_{\widetilde{q}}$~\footnote{Since
equation (\ref{eq:squark}) is a 1-loop approximation to the 2-loop RGEs and also contains a small constant term, and the plots have a
limited precision due to the 20-GeV step size in $m_{1/2}$ and $m_0$, 
small deviations in the $m_0$ and $m_{1/2}$ intercepts of the squark mass contours 
are expected.}. Since $C_{\widetilde{q}}$ decreases as $M_{in}$ decreases,
the semiminor axes of the squark mass contours increase progressively
between the panels of Fig.~\ref{fig:planes} and \ref{fig:moreplanes}.

We also show as the nearly vertical (green) lines in Figs.~\ref{fig:planes} and \ref{fig:moreplanes}
gluino mass contours from 0.5 - 3~TeV in 0.5~TeV increments.

\section{LHC Reach for Sparticle Discovery}

The discovery potential of ATLAS was examined  
in~\cite{tovey}, and more recently a CMS analysis~\cite{cmstdr} has
provided reach contours in the $(m_0,m_{1/2})$ plane within the CMSSM for $\tbt = 10$. 
These contours represent the total sensitivity to all the dominant processes expected to
occur at the LHC. To a good approximation, the contours depend only
on $m_{\widetilde{q}}$ and $m_{\widetilde{g}}$, although processes involving other 
gauginos and sleptons may become important near the focus-point and coannihilation 
strips~\cite{tovey2}. A full analysis of all the
processes involved in the estimation of the reach contours is beyond the
scope of this work, so we simply express the reach
contours as functions of $m_{\widetilde{q}}$ and $m_{\widetilde{g}}$,
and examine how the approximated reach in the $(m_{1/2}, m_0)$
plane changes as a function of $M_{in}$.

\begin{figure}
\begin{center}
\epsfig{file=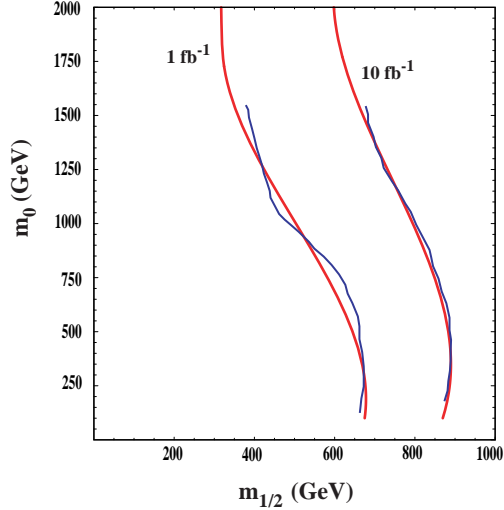,height=7cm}
\end{center}
\caption{\it Approximate LHC supersymmetry reach contours for integrated luminosities
of 1~fb$^{-1}$ and 10~fb$^{-1}$ (smooth curves), compared with the expected CMS reach given
in the CMS TDR~\protect\cite{cmstdr} for $\tbt = 10$.
\label{fig:approxreach}}
\end{figure}

We start with the 5-$\sigma$ inclusive supersymmetry discovery
contours in the CMSSM for 1.0 and 10~fb$^{-1}$ of integrated LHC luminosity,
shown for $\tbt = 10$ in Figure 13.5 of Ref.~\cite{cmstdr}. Since the inclusive reach is
expected to be fairly linear above $m_0 = 1.5$~TeV, we extend these
contours linearly above $m_0 = 1200$~GeV, then fit the sensitivity with
a third-order polynomial in $m_0$ and $m_{1/2}$ to extend the approximate LHC
supersymmetry reach out to $m_0 = 2$~TeV, as shown in Fig.~\ref{fig:approxreach}. 
Our fits are compared with the CMS reaches in Fig.~\ref{fig:approxreach}.
The largest differences between our approximate reach contours
and the contours shown in the CMS TDR~\cite{cmstdr} are $\sim 25$~GeV
for the 10~fb$^{-1}$ contour and $\sim 50$~GeV for the 1~fb$^{-1}$ contour. 

The next step is to change variables from $(m_{1/2},m_0) \goto
(m_{\widetilde{g}},m_{\widetilde{q}})$ using (\ref{eq:gauginos}) and (\ref{eq:squark}). 
Starting from the contours specified in Fig.~\ref{fig:approxreach} as
functions of the gluino and squark masses, for each value of $M_{in}$,
we then translate the discovery contours back into the corresponding
$(m_{1/2},m_0)$ plane. Clearly, the contours move as the universality scale 
$M_{in}$ is lowered and the gluino and squark masses change according to
(\ref{eq:gauginos}) and (\ref{eq:squark}).

The approximate 5-$\sigma$ discovery potential contours for the LHC with 1.0 and
10 fb$^{-1}$ of integrated luminosity are superposed as dashed (solid) dark blue lines
in the $(m_{1/2},m_0)$ planes for $\tbt = 10$, $A_0 = 0$ and different
values of $M_{in}$ in Figs.~\ref{fig:planes} and \ref{fig:moreplanes}.
We recall that the squark and gluino mass contours in the 
range (500, 3000)~GeV in
increments of 500 GeV are also shown, and that the squark
and gluino contours move to larger $m_{1/2}$ as $M_{in}$ is lowered,
resulting in more of the plane being accessible at a given luminosity.

We are unaware of any up-to-date study of the regions of the $(m_{1/2},m_0)$ plane
that could be excluded at the 95~\% C.L. by the LHC with a specified integrated luminosity. However,
in previous studies the 95~\% exclusion reach was similar to the 5-$\sigma$ discovery with a
factor $\sim 5$ more luminosity. Therefore, we estimate that the `discovery' regions 
of Figs.~\ref{fig:planes} and \ref{fig:moreplanes} bounded by the (dark blue) dashed and solid
lines could, alternatively, be excluded by the LHC with a factor of $\sim 5$ less luminosity, namely 
$\sim 0.2 (2)$~fb$^{-1}$.

\section{Impact of the Cold Dark Matter Density Constraint}

We now consider the consequences if the relic neutralino LSP density lies within the range
\beq
0.088 \; < \; \Omega_{\chi} h^2 \; < \; 0.12
\eeq
favoured by WMAP and other astrophysical and cosmological measurements~\cite{WMAP}.
In  Figs.~\ref{fig:planes} and \ref{fig:moreplanes}, 
the corresponding strips of preferred density in
the various $(m_{1/2}, m_0)$ planes are shaded (light turquoise). In the case of the
CMSSM, shown in panel (a), the coannihilation strip extends up to $(m_{1/2}, m_0)
\sim (900, 220)$~GeV, and it all lies within the LHC supersymmetry discovery reach,
which extends to $m_{\tilde g} = 2000$~GeV with
$10$~fb$^{-1}$ of integrated luminosity, also corresponding to $(m_{1/2}, m_0)
\sim (900, 220)$~GeV. Moreover, the underdense region lying between
the WMAP coannihilation strip and the boundary of the (brown shaded) charged-LSP region, 
where $\Omega_{\chi} h^2 < 0.088$, is also accessible
to the LHC. However, the WMAP strip in the focus-point region, and the corresponding
underdense region lying between it and the (pink shaded) electroweak symmetry-breaking 
boundary is only partially accessible to the LHC. For this reason, there is no `guarantee' of finding
supersymmetry at the LHC, even within the CMSSM at this value of $\tan \beta$.

Turning now to GUT-less models,
the full coannihilation strip and the corresponding underdense region are also fully
accessible to the LHC for $M_{in} = 10^{14}$~GeV as the endpoint of the 
coannihilation strip moves to smaller $m_{1/2}$, as seen in panel (b) of
Fig.~\ref{fig:planes}. However, when $M_{in} = 10^{13}$~GeV, as shown in 
panel (c) of Fig.~\ref{fig:planes}, the coannihilation strip merges into a rapid-annihilation
funnel that does not appear in the CMSSM for
this value of $\tbt = 10$. To its right there is another very narrow WMAP-compatible strip and, at even
larger $m_{1/2}$, an overdense region extending (almost) to the boundary of the (brown shaded)
forbidden charged-LSP region.Whilst a substantial portion of the $(m_{1/2}, m_0)$
plane will be probed at the LHC, there are now regions of both WMAP-compatible
regions (focus-point and coannihilation/funnel) that are inaccessible to the LHC. Moreover, 
there are now also large underdense regions at large $m_{1/2}$ and $m_0$, above the preferred
focus-point strip and to the right of the coannihilation strip, that are also inaccessible to the LHC. 

When $M_{in}$ is reduced to $10^{12.5}$~GeV, as seen in panel (d) of
Fig.~\ref{fig:planes}, the focus-point and coannihilation strips join to form an `atoll'. Inside its
`lagoon', the relic density is in general too large, whereas the region around the `atoll' is underdense.
At larger values of $M_{1/2}$ than the `atoll', there is a narrow strip that is the vestige of the
other side of the rapid-annihilation funnel, beyond which the relic density is again too 
large\footnote{The chain of small `islands' seen within 
the `atoll' are caused by the $s$-channel coannihilation of  $\chi_1 \chi_2$
through heavy Higgs scalars and pseudoscalars, which brings the relic density down into
the WMAP range along a very narrow neutralino coannihilation funnel.  This is seen as 
a string of `islands' rather than as a  `peninsula'
because of the finite resolution of our scan.}.
The LHC provides access to a significant fraction of the `atoll' and the surrounding
underdense region, but only a small part of the strip beyond the funnel.

When $M_{in}$ is reduced to $10^{12}$~GeV, as seen in panel (a) of
Fig.~\ref{fig:moreplanes}, the `atoll' contracts to a WMAP-compatible `island' centred around 
$(m_{1/2}, m_0) \sim (600, 700)$~GeV that is completely accessible to the LHC with an
integrated luminosity of 10~fb$^{-1}$. There is also a 
WMAP-compatible `mark of Zorro' extending to larger $m_{1/2}$ that is only partially accessible to
the LHC. Its narrow diagonal is due to the crossing of the $h A$ threshold in LSP annihilations. The large region surrounding the `island' is underdense, and
accessible only partially to the LHC. The relic density is too high in the region below
the `mark of Zorro', and beyond it falls below the WMAP range.

When $M_{in}$ is further decreased to $10^{11.5}$~GeV, as seen in panel (b) 
of Fig.~\ref{fig:moreplanes},  the relic density is WMAP-compatible only along a strip
close to the boundary of the stau LSP region. The LHC still has some chance of detecting 
sparticles in the cold dark matter region in this case, since the WMAP-compatible
strip starts at $m_{1/2} \sim 600$~GeV. 

However, the situation changes dramatically in the case
$M_{in} = 10^{11}$~GeV, shown in panel (c) of Fig.~\ref{fig:moreplanes}. In this case, the only
WMAP-compatible region is a small ellipsoid at $(m_{1/2}, m_0) \sim
(2000, 1100)$~GeV, beyond the reach of the LHC, which is surrounded by an only
partially-accessible underdense region of the $(m_{1/2}, m_0)$ plane. The WMAP-compatible
region is similar for $M_{in} = 10^{10}$~GeV, as shown in panel (d) of Fig.~\ref{fig:moreplanes}.

We conclude that the prospects for discovering supersymmetry at the LHC in scenarios
where the neutralino LSP provides some of the cold dark matter are in general {\it diminished}
in GUT-less scenarios. In particular, the `guarantee' that the LHC would find supersymmetry 
if $\tbt = 10$, which was valid in the coannihilation region of the CMSSM but not in the focus-point
region, is not valid in GUT-less models. Indeed, if $M_{in} < 10^{11.5}$~GeV, the LHC provides
access to {\it none} of the WMAP-preferred region.

\section{Sparticle Pair Production at Linear $e^+ e^-$ Colliders}
\label{sec:pairprod}

In this Section, we examine the sparticle pair production threshold in $e^+ e^-$
collisions in light of the above discussion. The
area of the $(m_{1/2},m_0)$ plane accessible to ATLAS and CMS clearly
{\it increases} as the integrated LHC luminosity {\it increases}, and also (slightly,
as we have already noted)
as $M_{in}$ {\it decreases}. Here we ask the following questions\footnote{These
questions were raised previously in the CMSSM context in~\cite{LHC}.}:
if a signature of new physics
is observed at a given luminosity, what is the $e^+ e^-$ centre-of-mass energy 
at which sparticles are {\it guaranteed} to be pair produced and, conversely, if no
sparticles have (yet) been seen at the LHC, what is the $e^+ e^-$ centre-of-mass energy 
at which sparticles are guaranteed {\it not} to be pair produced? We continue to
focus on the case $\tbt = 10$ and $A_0 =0$, since this is the value for which up-to-date LHC discovery 
contours are available\footnote{The cosmologically preferred regions in the
$(m_{1/2}, m_0)$ planes depend in general on $A_0$ as well as
$\tbt$.}. Also, in the following discussion we focus on $M_{in} \ge 10^{11.5}$~GeV, 
since the LHC does not provide access to any of the WMAP-preferred region for lower
values of $M_{in}$.

\begin{figure}[p]
\begin{center}
\mbox{\epsfig{file=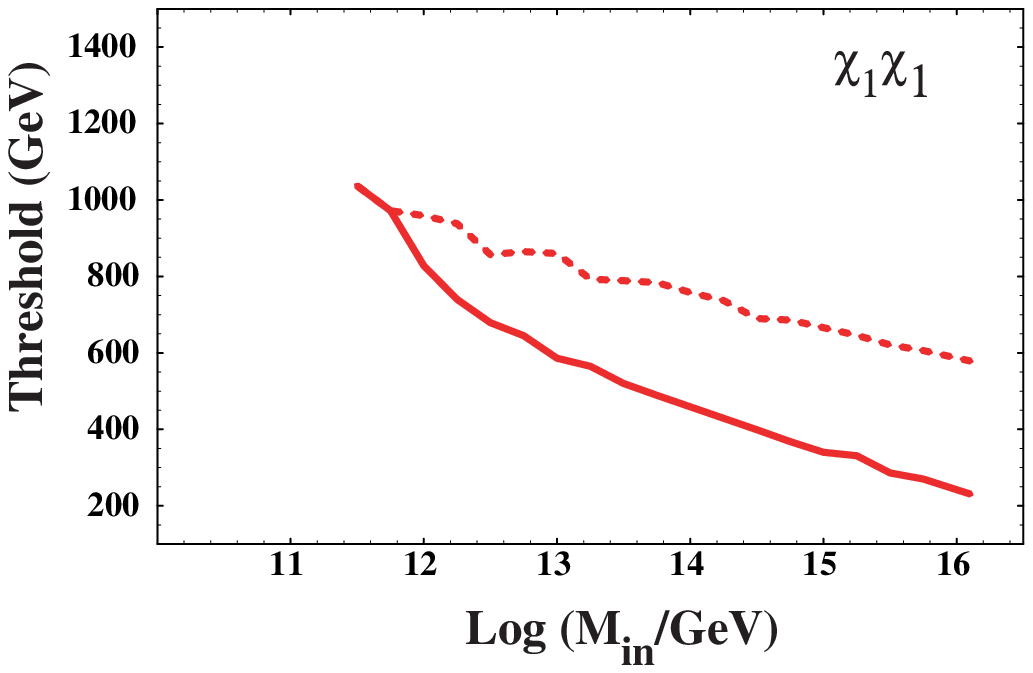,height=7cm}}
\end{center}
\begin{center}
\mbox{\epsfig{file=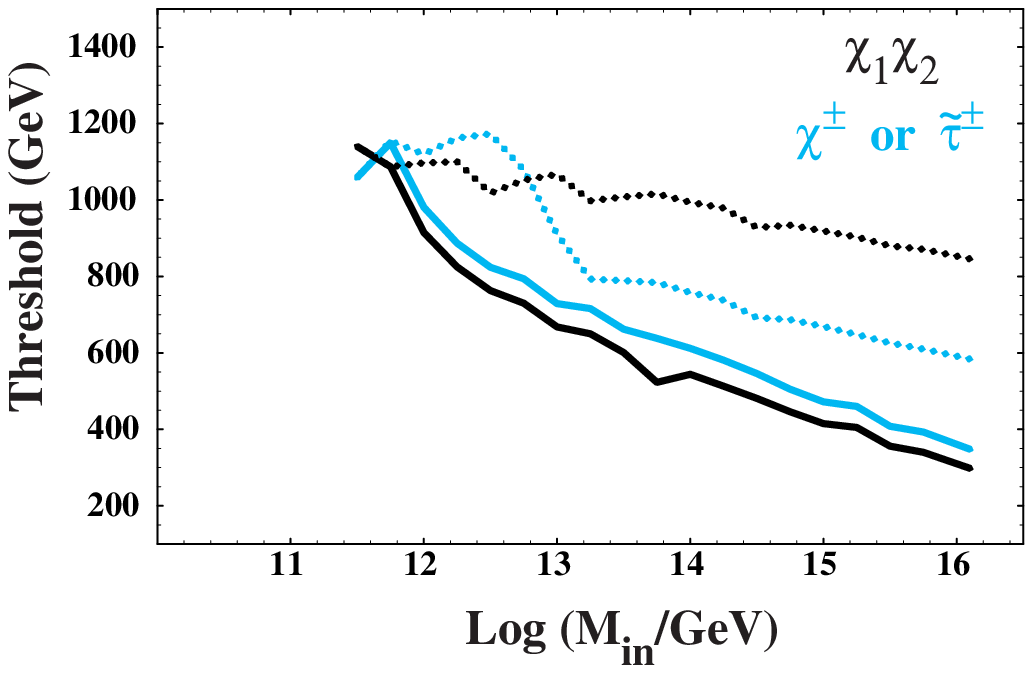,height=7cm}}
\end{center}
\caption{\it Pair-production $e^+ e^-$ thresholds for the lightest neutralinos are shown in panel 
(a), and the thresholds
for charged-sparticle pair production (light blue) and associated $\chi_0^1 \chi_0^2$ production 
(black) are shown in 
panel (b). The dashed curves show the $e^+ e^-$ centre-of-mass energy required for a `guarantee' that
the corresponding sparticles can be produced at a LC, if supersymmetry is discovered at the LHC with
1.0 fb$^{-1}$ of data. The solid lines give the lower limit on the thresholds if the LHC
establishes that there is no supersymmetry within this discovery reach. We
assume that the cold dark matter density falls
within the range favoured by WMAP and that $\tbt = 10$ and $A_0 = 0$.
\label{fig:ILC1}}
\end{figure}

In the scenarios considered here, the LSP is the lightest neutralino $\chi$,
so a linear $e^+ e^-$ collider will pair-produce sparticles if the
centre-of-mass energy $E_{CM} > 2 m_\chi$. With sufficient luminosity,
the radiative reaction $e^+ e^- \to \chi \chi \gamma$ may be detectable quite
close to the pair-production threshold. Failing this, along the coannihilation
strip close to the kinematic boundary where $m_\chi = m_{\tilde \tau_1}$,
one expects only a small mass difference $m_{\tilde \tau_1} - m_\chi$, so that
the threshold for $e^+ e^- \to {\tilde \tau_1}^+ {\tilde \tau_1}^-$ production
and detection would lie only slightly above the $e^+ e^- \to \chi \chi$ threshold.
Other processes that should be detectable include $e^+ e^- \to \chi \chi_2$
and $e^+ e^- \to \chi^+ \chi^-$. In the following, we consider all these processes.
We display in Figs.~\ref{fig:ILC1} and \ref{fig:ILC10} pair-production thresholds
for these different $e^+ e^-$ reactions as functions of $M_{in}$, assuming that
the relic density of neutralinos falls within the range
$ 0.088 < \Omega_{\chi} h^2 < 0.12 $ preferred by WMAP and others.

\subsection{The CMSSM Case: $M_{in} = M_{GUT}$}

We consider first the sparticle production thresholds corresponding
to an LHC luminosity of 1.0 fb$^{-1}$, which are shown in
Fig.~\ref{fig:ILC1}.  In the usual GUT-scale CMSSM, the LHC 
1.0~fb$^{-1}$ discovery contour crosses a cosmologically-preferred region of the
$(m_{1/2},m_0)$ plane in two places, as one can see from panel (a) of
Fig.~\ref{fig:planes}. One crossing occurs in the focus-point region,
at approximately $(300,1530)$~GeV. Here, the lightest neutralino is a mixed state,
with $m_{\widetilde{\chi}} = 115$~GeV. The other crossing
of the 1.0~fb$^{-1}$ LHC contour with a cosmologically-preferred
region occurs along the coannihilation strip, which borders the
excluded $\stau$-LSP region at low $m_0$.  This crossing occurs at
$(680,160)$~GeV.  Since $m_{1/2}$ is larger here, the neutralino LSP
is correspondingly heavier, with $m_{\widetilde{\chi}} = 290$~GeV.
We conclude that, if $M_{in} = M_{GUT}$ and sparticles are discovered at the
LHC with 1.0~fb$^{-1}$ of data, then neutralino LSP pairs
would definitely be produced at a linear collider with a centre-of-mass
energy $E_{cm} = 580$~GeV or more. This threshold is displayed as the starting point at $M_{in} = M_{GUT}$ of the dashed line in the upper panel of Fig.~\ref{fig:ILC1}. Conversely, if the LHC
establishes that supersymmetry does not exist in this 1.0~fb$^{-1}$ discovery 
region\footnote{We recall that this conclusion might be possible with an analysis of
0.2~fb$^{-1}$ of integrated luminosity.}, the LSP
must weigh at least 115~GeV, and hence the LC threshold for $\chi \chi$ production must be
at least 230~GeV, which is the starting point of the solid line in the upper panel of Fig.~\ref{fig:ILC1}.

Charginos are also relatively light in the focus-point
region, whereas the sfermions are all much heavier\footnote{This is why this region
is disfavoured by the experimental range of $g_\mu - 2$: see the pink shaded
region in panel (a) of Fig.~\ref{fig:planes}.}.  For example, at the point in the focus-point
region where the LHC 1.0~fb$^{-1}$ discovery curve crosses the WMAP strip, the
chargino (which has a large Higgsino component) weighs 175~GeV, whereas the
lighter stau has $m_{\stau_1} = 1520$~GeV. The lighter stop and sbottom
squarks are somewhat lighter, with $m_{\widetilde{t}_1} = 1035$~GeV
and $m_{\widetilde{b}_1} = 1350$~GeV.  On the other hand, at the intersection
of the LHC 1.0~fb$^{-1}$ discovery curve with the WMAP strip in the coannihilation
region, the mass of the lighter stau is very similar to that of the LSP, at $m_{\stau} = 292$~GeV. 
The right-handed selectron and smuon are also light in this case, but most sfermions are
considerably heavier with masses in the TeV range: the lighter
chargino is gaugino-dominated, with $m_{\widetilde{\chi}^\pm} = 555$~GeV. 
The corresponding thresholds for charged-sparticle pair production
are displayed as the starting points at $M_{in} = M_{GUT}$
of the lighter (blue) dashed and solid lines in the lower panel of Fig.~\ref{fig:ILC1}. The dashed
line represents the centre-of-mass energy $\sim 585$~GeV that a LC would need for a `guarantee' of
producing charged-sparticle pairs if the LHC discovers supersymmetry with 1.0~fb$^{-1}$, and the
solid line represents the lowest centre-of-mass energy $\sim 350$~GeV
where they might still appear at a LC even if
the LHC excludes this 1.0~fb$^{-1}$ discovery region.

The thresholds for associated $\chi \chi_2$ production are in general
intermediate between the $\chi \chi$ and $\chi^+ \chi^-$ thresholds, since
$m_{\chi_2} \sim m_{\chi^\pm}$. Thus, the starting points at $M_{in} = M_{GUT}$
of the $\chi \chi_2$ threshold lines, shown as the darker (black) lines
in the lower panel of Fig.~\ref{fig:ILC1}, are lower than those for $\chi^+ \chi^-$ in the
focus-point region ($E_{cm} = 290$~GeV, starting point of the solid line) and higher than that for 
${\tilde \tau_1}^+ {\tilde \tau_1}^-$ production in the coannihilation region
($E_{cm} = 845$~GeV, starting point of the dashed line). Again, $E_{cm}$ above the
dashed line would `guarantee' $\chi \chi_2$ at a LC if the LHC discovers supersymmetry with 
1.0~fb$^{-1}$, whereas the threshold must lie above the solid line if the LHC in fact excludes
the existence of supersymmetry within this discovery region.

In summary: if the LHC discovers sparticles with an integrated luminosity
of 1.0~fb$^{-1}$, a centre-of-mass energy $\sim 600 (850)$~GeV would be required
for a LC to be `guaranteed' to pair-produce LSPs and charged sparticles ($\chi \chi_2$)
within the CMSSM framework. On the other hand, within the CMSSM, the corresponding LC
thresholds would be $\gappeq 230$ and 350 (290)~GeV if the LHC in fact excludes
supersymmetry within the 1.0~fb$^{-1}$ discovery region.

\subsection{The GUT-less Case: $M_{in} < M_{GUT}$}

As the assumed scale of universality of the soft supersymmetry-breaking
parameters is reduced from the supersymmetric GUT scale of $M_{GUT} \sim
2 \times 10^{16}$ GeV, the sparticle masses evolve as
exemplified in Fig.~1. Correspondingly, the inclusive LHC sparticle
reach in the $(m_{1/2}, m_0)$ plane changes as discussed in Section~3. 
In addition, the cosmologically-preferred
regions in the $(m_{1/2}, m_0)$ plane also move, as described in depth in~\cite{eos1,eos2}, 
and as seen in Figs.~2 and 3 and discussed in Section~4. Consequently, the LC
thresholds discussed in the previous subsection also change, as seen in 
Fig.~\ref{fig:ILC1}, which we now discuss in more detail.

In general, as already discussed, the renormalizations of the sparticle masses are
reduced and the sparticle spectrum is correspondingly compressed as $M_{in}$
decreases. As a result, as seen from the dashed line in the upper panel of Fig.~\ref{fig:ILC1}, the LC
centre-of-mass energy corresponding to a given LHC reach generally
{\it increases} as $M_{in}$ {\it decreases}. As $M_{in}$ varies, the LHC discovery contour may
intersect the WMAP-preferred region in more than two places (see, e.g., panel (d) of
Fig.~\ref{fig:planes} for $M_{in} = 10^{12.5}$~GeV), or even in a continuum of points 
(see, e.g., panel (a) of Fig.~\ref{fig:moreplanes} for $M_{in} = 10^{12}$~GeV). Here and in the
following discussion, the dashed lines always correspond to the largest value that the
corresponding threshold can take at any of these points, and the solid lines
correspond to the smallest of these values. Thus, the dashed lines represent the $E_{cm}$
above which sparticle production is `guaranteed' at a LC if the LHC discovers supersymmetry
with 1~fb$^{-1}$ of data, and the solid lines represent the minimum value that the
threshold could have if this region is excluded.

In the upper panel of Fig.~\ref{fig:ILC1}, the dashed line rises fairly steadily as $M_{in}$
decreases. The slight flattening between $\log M_{in} = 13.3 - 13.7$ 
is because the LHC discovery reach
extends beyond the tip of the coannihilation strip. However, when $M_{in} \lappeq
10^{13.3}$~GeV, the coannihilation strip sprouts a rapid-annihilation funnel (see panel (c) of
Fig.~\ref{fig:planes}), and the maximum possible value of $m_\chi$ increases again.
The irregularities visible in the dashed lines in the lower panel of Fig.~\ref{fig:ILC1} have
similar origins.
For $M_{in} \lappeq10^{11.8}$~GeV, the LHC discovery contour meets the WMAP-preferred
region in just one location (see panel (b) of Fig.~\ref{fig:moreplanes}, 
and the dashed and solid lines merge, as seen in both panels of Fig.~\ref{fig:ILC1}.
We recall that there is no LHC-accessible region for $M_{in} < 10^{11.5}$~GeV, so
both the dashed and solid lines are truncated at this value.
In order to `guarantee' pair-production of LSPs,
whatever the value of $M_{in} > 10^{11.5}$~GeV, a LC with $E_{cm} > 1040$~GeV
would be required if the LHC discovers supersymmetry with 1~fb$^{-1}$ of data. 
For $M_{in} = 10^{11.5}$~GeV, a similar $E_{cm}$ would be required
for a LC to have any chance of producing $\chi$ pairs if the LHC actually excluded this
1~fb$^{-1}$ discovery region. However, the solid line shows that smaller $E_{cm}$ might 
be sufficient if $M_{in}$ is larger.

Analogous effects as $M_{in}$ decreases are seen for charged-sparticle
pair production, as shown by the lighter solid and dashed lines in the lower
panel of Fig.~\ref{fig:ILC1}. There is, however, a complication induced by the
fact that one should keep in mind several different charged-sparticle masses,
principally $m_{\tilde \tau_1}$ and $m_{\chi^\pm}$. In general, the light (blue)
dashed line represents the upper limit on the lowest charged-sparticle threshold, and the
light (blue) solid line represents the lower limit on the lowest charged-sparticle threshold.
As in the LSP case shown in the upper panel of Fig.~\ref{fig:ILC1},
the dashed and solid lines merge when $M_{in} < 10^{12}$~GeV.  Overall, in order 
to `guarantee' charged-sparticle pair production, whatever the value of $M_{in}$,
a LC with $E_{cm} > 1180$~GeV would be required.

Finally, we consider the example of associated $\chi \chi_2$ production, shown as
the darker (black) solid and dashed lines in the lower panel of Fig.~\ref{fig:ILC1}.
As previously, the threshold required for a `guarantee' tends to
increase as $M_{in}$ decreases, and a LC with
$E_{cm} > 1140$~GeV would be required to `guarantee' the observability
of associated $\chi \chi_2$ production.

\subsection{Integrated LHC Luminosity of 10~fb$^{-1}$}

\begin{figure}[p]
\begin{center}
\mbox{\epsfig{file=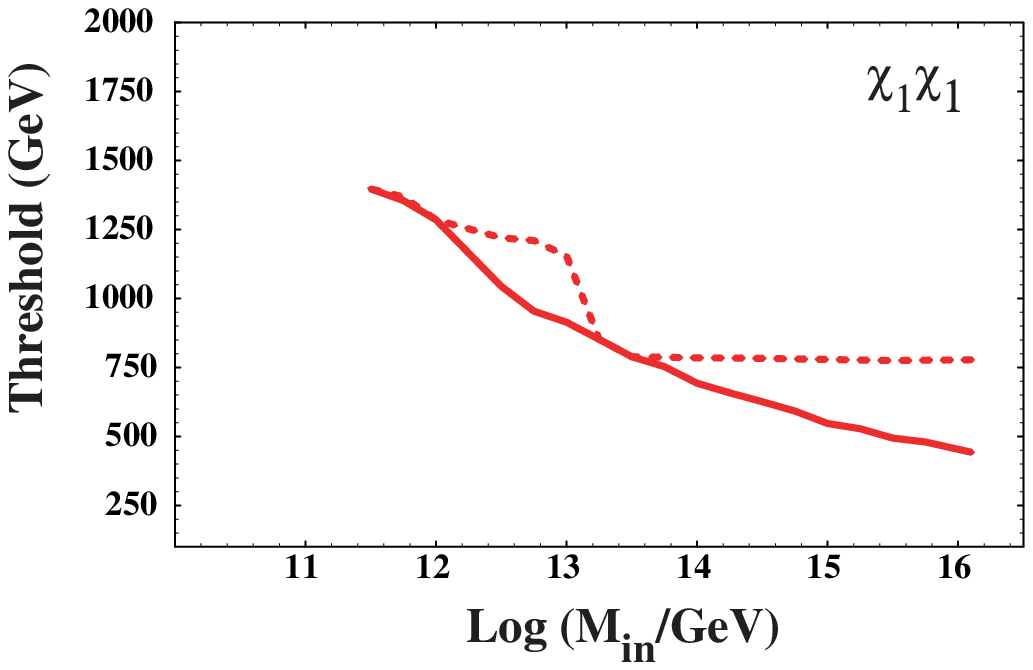,height=7cm}}
\end{center}
\begin{center}
\mbox{\epsfig{file=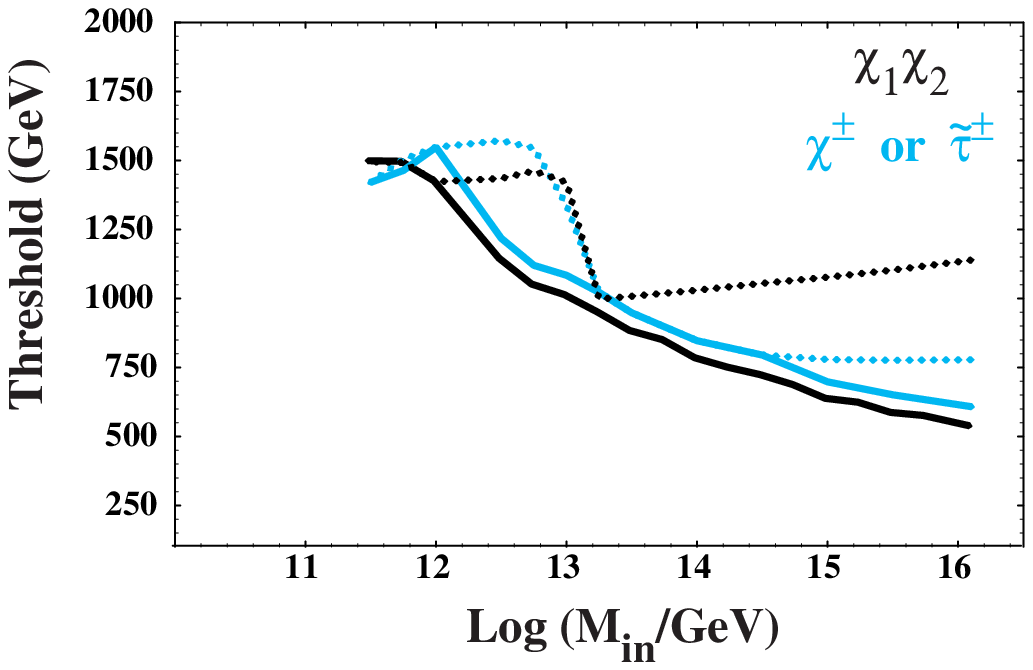,height=7cm}}
\end{center}
\caption{\it As for Fig.~\protect\ref{fig:ILC1}, assuming that
the LHC discovers supersymmetry with 10 fb$^{-1}$ of data,
or excludes it within this discovery reach.
\label{fig:ILC10}}
\end{figure}

A similar story unfolds for an LHC integrated luminosity of 
10~fb$^{-1}$, as shown in Fig.~\ref{fig:ILC10}. While the general
behaviour of the thresholds as a function of $M_{in}$ is roughly the
same, there are two important differences.  First, the thresholds are
in general larger. In the case of an LHC discovery, the upper limits on the
sparticle pair-production thresholds are typically about 30-35 \% larger for
$M_{in} = M_{GUT}$. However, a second difference is that the
coannihilation strip is now contained within the LHC discovery reach
for $10^{13.3}$~GeV $< M_{in} \le M_{GUT}$, implying that the corresponding range of
$m_{1/2}$ is unrelated to the accessible value of $m_{\tilde g}$. This leads to a plateau
in the $\chi \chi$ `guarantee' threshold and even a decrease in the
$\chi \chi_2$ `guarantee' threshold as $M_{in}$ decreases over this range.
In fact, the curves even
merge near $M_{in}=10^{13.3}$ GeV, where the heaviest $\stau_1$
in the coannihilation strip is lighter than the lightest $\chi^{\pm}$
from the focus point. At this point, the energy required to `guarantee'
that charged sparticles are pair produced is the $\chi^+\chi^-$
threshold, which, since the coannihilation strip terminates inside the
LHC reach contour, is also the minimum energy at which pair production
could be expected if the area inside that contour is excluded.

Looking at the upper limits on the threshold for $\chi \chi$
pair production shown in the upper panel of Fig.~\ref{fig:ILC10},
we see that the values for $M_{in} = M_{GUT}$ are significantly larger
than for the case of 1~fb$^{-1}$ shown in Fig.~\ref{fig:ILC1}, reflecting the
improved physics reach of the LHC with 10~fb$^{-1}$. A
centre-of-mass energy of at least 800~GeV would be required to `guarantee'
$\chi \chi$ production for large $M_{in}$, increasing to 1.4~TeV for $M_{in} \sim 10^{11.5}$~GeV
(see the dashed red line).
Conversely, the absence of supersymmetry within the LHC 10~fb$^{-1}$
discovery region\footnote{We recall that this is a possible outcome with $\sim 2$~fb$^{-1}$ of
analyzed LHC data.} would imply (see the solid red line) that the LC threshold for
$\chi \chi$ production must be at least 450~GeV for $M_{in} = M_{GUT}$, rising to 1.4~TeV for
for $M_{in} \sim 10^{11.5}$~GeV.

In the case of charged-sparticle pair production, shown 
as the lighter lines in the lower panel of
Fig.~\ref{fig:ILC10}, almost the same energy $\sim 800$~GeV would be
required to `guarantee' being
above threshold if $M_{in} = M_{GUT}$ (see the dashed light-blue line), 
whereas a LC with $E_{cm} > 1.6$~TeV 
would be required to `guarantee' the observability
of charged-sparticle pair production, whatever the value of $M_{in} > 10^{11.5}$~GeV.
Conversely, the absence of supersymmetry within the LHC 10~fb$^{-1}$
discovery region would imply (see the solid light-blue line) that the LC threshold for
charged-sparticle pair production must be at least 600~GeV for $M_{in} = M_{GUT}$, rising to 
1.5~TeV for $M_{in} \sim 10^{11.5}$~GeV.

Finally, in the case of associated $\chi \chi_2$ production, shown 
as the darker lines in the lower panel of Fig.~\ref{fig:ILC10}, the energy
required to `guarantee' being above threshold  is
$\gtrsim 1.1$~TeV for $M_{in} = M_{GUT}$ (see the dashed black line), 
decreasing somewhat to $\sim 950$~GeV for
\mbox{$M_{in} \sim 10^{13.5}$~GeV}.
On the other hand, the absence of supersymmetry within the LHC 10~fb$^{-1}$
discovery region would imply (see the solid black line) that the LC threshold for
$\chi \chi_2$ pair production must be at least 550~GeV for $M_{in} = M_{GUT}$, rising 
monotonically to 1.5~TeV for $M_{in} \sim 10^{11.5}$~GeV.

\section{Conclusions}

We have discussed in the previous Section how much centre-of-mass energy
would be required to `guarantee' the observability of sparticle pair production
in $e^+ e^-$ collisions under various hypotheses for the integrated luminosity
required for discovering supersymmetry at the LHC and for different values of the
universality scale $M_{in}$. We have also discussed how corresponding
sparticle exclusions at the LHC would set lower limits on the possible
thresholds for producing different sparticle pairs at a LC.
To conclude, we now consider the capabilities of LCs with various specific
proposed centre-of-mass energies.

Even if supersymmetry were to be 
found at the LHC with 1~fb$^{-1}$ of integrated luminosity, 
a LC with $E_{cm} = 0.5$~TeV would not be `guaranteed' to produce
$\chi \chi$ pairs or other sparticle pairs.
However, even if supersymmetry were to be excluded in the LHC's 1~fb$^{-1}$
discovery region, the possibility of observing sparticles at a LC with $E_{cm} = 0.5$~TeV could
not be excluded for $M_{in} > 10^{13.5}$~GeV, and such a LC might also
pair-produce charged sparticles if $M_{in} > 10^{15}$~GeV and/or
produce $\chi \chi_2$ in association if $M_{in} > 10^{14.5}$~GeV.
On the other hand, if supersymmetry were not even within the 10~fb$^{-1}$ 
discovery reach of the LHC, a LC with $E_{cm} = 0.5$~TeV might be (barely)
above the $\chi \chi$ threshold only if $M_{in} \gtrsim 10^{15.5}$~GeV, and there
would be no likelihood of charged-sparticle or $\chi \chi_2$ production.

A LC with $E_{cm} = 1$~TeV would be `guaranteed', if supersymmetry were to be 
found at the LHC with 1~fb$^{-1}$ of integrated luminosity, to produce
$\chi \chi$ pairs in any GUT-less scenario with $M_{in} > 10^{12}$~GeV.
Analogous `guarantees' for charged-sparticle pair production or associated
$\chi \chi_2$ production could be given only for $M_{in} > 10^{13} (10^{14})$~GeV,
respectively. On the other hand, if supersymmetry were not even within the 10~fb$^{-1}$ 
discovery reach of the LHC, it might still be possible to find $\chi \chi$ (charged-sparticle pairs)
($\chi \chi_2$) at a LC if $M_{in} > 10^{12.5} (10^{13.3}) (10^{13})$~GeV.

Finally, even if the LHC would require 10~fb$^{-1}$ to discover supersymmetry,
a LC with $E_{cm} = 1.5$~TeV would be `guaranteed' to produce $\chi \chi$
and $\chi \chi_2$ pairs in all the allowed WMAP-compatible scenarios, and
charged-sparticle pair production would be `guaranteed' for all except a small range of
$M_{in}$ between $10^{12}$ and $10^{13}$~GeV. Hence, a LC with
$E_{cm} = 1.5$~TeV would be well matched to the physics reach of the LHC
with this luminosity, whereas
a LC with a lower $E_{cm}$ might well be unable to follow up on a discovery of
supersymmetry at the LHC. However, as already mentioned, even in the absence of
any `guarantee', it could still be that
the LHC discovers supersymmetry at some mass scale well below the limit of its
sensitivity with 10~fb$^{-1}$ of integrated luminosity, in which case a lower-energy
LC might still have interesting capabilities to follow up on a discovery of 
supersymmetry at the LHC.

It is clear that the physics discoveries of the LHC will be crucial for the scientific
prospects of any future LC. Supersymmetry is just one of the scenarios whose
prospects at a LC may depend on what is found at the LHC. Even within the
supersymmetric framework, there are many variants that should
be considered. Even if $R$ parity is conserved, the LSP might not be the lightest
neutralino. Even if it is, the relevant supersymmetric model may not be
minimal. Even if it is the MSSM, supersymmetry breaking may not be universal. Even
if it is, the universality scale may not be the same for gauginos and sfermions.
Nevertheless, we hope that study serves a useful purpose in highlighting some
of the issues that may arise in guessing the LC physics prospects on the basis
of LHC physics results.

\section*{Acknowledgments}
\noindent 
We would like to thank Dan Tovey and Maria Spiropulu for useful
conversations and information.
The work of K.A.O. and P.S. was supported in part
by DOE grant DE--FG02--94ER--40823.


\begin{thebibliography}{99}

\bibitem{lhc}
ATLAS Collaboration, 
  {\em Detector and Physics Performance Technical Design Report},
  CERN/LHCC/99-15 (1999), see:\\
  {\tt http://atlasinfo.cern.ch/Atlas/GROUPS/PHYSICS/TDR/access.html}~;
 M.~Schumacher,
  {\em Czech. J. Phys.} {\bf 54} (2004) A103;
  arXiv:hep-ph/0410112;
 S.~Abdullin {\it et al.},
              {\em Eur. Phys. J.} {\bf C 39S2} (2005) 41.
              
\bibitem{cmstdr}
  The CMS Collaboration,
  {\it CMS Physics Technical Design Report. Volume II: Physics
   Performance}, CERN/LHCC 2006-021, CMS TDR 8.2 (2006),
    see:
                  {\tt http://cmsdoc.cern.ch/cms/cpt/tdr/}~.

\bibitem{LHC}
   J.-J.~Blaising, A.~De Roeck, J.~Ellis, F.~Gianotti, P.~Janot,
     G.~Rolandi, and D.~Schlatter,
  {\it Potential LHC Contributions to Europe's Future Strategy at the
    High Energy Frontier};
   J.~Ellis,
    arXiv:hep-ph/0611237.  


\bibitem{ilc}
J.~Aguilar-Saavedra {\it et al.},
                   TESLA TDR Part~3: 
                   {\it Physics at an $e^+e^-$ Linear Collider},
                   arXiv:hep-ph/0106315,
                   see: {\tt http://tesla.desy.de/tdr/} ;
T.~Abe {\it et al.}
                     [American Linear Collider Working Group Collaboration],
                     {\it Resource book for Snowmass 2001}, 
                     arXiv:hep-ex/0106055;
K.~Abe {\it et al.} 
                  [ACFA Linear Collider Working Group Collaboration],
                  arXiv:hep-ph/0109166.
S.~Heinemeyer {\it et al.},
                          arXiv:hep-ph/0511332.
                          
 \bibitem{EHNOS} H.~Goldberg,
                {\em Phys. Rev. Lett.} {\bf 50} (1983) 1419;
                J.~Ellis, J.~Hagelin, D.~Nanopoulos, K.~Olive and M.~Srednicki,
                {\em Nucl. Phys.} {\bf B 238} (1984) 453.

\bibitem{mssm}H.~Nilles, 
               {\em Phys.\ Rept.} {\bf 110} (1984) 1;
H.~Haber and G.~Kane, 
               {\em Phys.\ Rept.} {\bf 117} (1985) 75; \\
               R.~Barbieri, 
               {\em Riv.\ Nuovo Cim.} {\bf 11} (1988) 1. 
               
\bibitem{Baer}
H.~Baer, C.~Balazs, A.~Belyaev, T.~Krupovnickas and X.~Tata,
JHEP {\bf 0306} (2003) 054
[arXiv:hep-ph/0304303].

\bibitem{funnel}
M.~Drees and M.~M.~Nojiri,
Phys.\ Rev.\ D {\bf 47} (1993) 376 [arXiv:hep-ph/9207234];
H.~Baer and M.~Brhlik,
Phys.\ Rev.\ D {\bf 53} (1996) 597 [arXiv:hep-ph/9508321];
  Phys.\ Rev.\  D {\bf 57} (1998) 567
  [arXiv:hep-ph/9706509];
H.~Baer, M.~Brhlik, M.~A.~Diaz, J.~Ferrandis, P.~Mercadante, P.~Quintana and X.~Tata,
  Phys.\ Rev.\  D {\bf 63} (2001) 015007
  [arXiv:hep-ph/0005027];
 A.~B.~Lahanas, D.~V.~Nanopoulos and V.~C.~Spanos,
  Mod.\ Phys.\ Lett.\  A {\bf 16} (2001) 1229
  [arXiv:hep-ph/0009065].

\bibitem{cmssm}
J.~R.~Ellis, T.~Falk, K.~A.~Olive and M.~Schmitt,
Phys.\ Lett.\ B {\bf 388} (1996) 97
[arXiv:hep-ph/9607292];
Phys.\ Lett.\ B {\bf 413} (1997) 355
[arXiv:hep-ph/9705444];
J.~R.~Ellis, T.~Falk, G.~Ganis, K.~A.~Olive and M.~Schmitt,
Phys.\ Rev.\ D {\bf 58} (1998) 095002
[arXiv:hep-ph/9801445];
V.~D.~Barger and C.~Kao,
Phys.\ Rev.\ D {\bf 57} (1998) 3131
[arXiv:hep-ph/9704403];
J.~R.~Ellis, T.~Falk, G.~Ganis and K.~A.~Olive,
Phys.\ Rev.\ D {\bf 62} (2000) 075010
[arXiv:hep-ph/0004169];
V.~D.~Barger and C.~Kao,
Phys.\ Lett.\ B {\bf 518} (2001) 117
[arXiv:hep-ph/0106189];
L.~Roszkowski, R.~Ruiz de Austri and T.~Nihei,
JHEP {\bf 0108} (2001) 024
[arXiv:hep-ph/0106334];
A.~B.~Lahanas and V.~C.~Spanos,
Eur.\ Phys.\ J.\ C {\bf 23} (2002) 185
[arXiv:hep-ph/0106345];
A.~Djouadi, M.~Drees and J.~L.~Kneur,
JHEP {\bf 0108} (2001) 055
[arXiv:hep-ph/0107316];
U.~Chattopadhyay, A.~Corsetti and P.~Nath,
Phys.\ Rev.\ D {\bf 66} (2002) 035003
[arXiv:hep-ph/0201001];
J.~R.~Ellis, K.~A.~Olive and Y.~Santoso,
New Jour.\ Phys.\  {\bf 4} (2002) 32
[arXiv:hep-ph/0202110];
H.~Baer, C.~Balazs, A.~Belyaev, J.~K.~Mizukoshi, X.~Tata and Y.~Wang,
JHEP {\bf 0207} (2002) 050
[arXiv:hep-ph/0205325];
R.~Arnowitt and B.~Dutta,
arXiv:hep-ph/0211417.

\bibitem{efgosi}
J.~R.~Ellis, T.~Falk, G.~Ganis, K.~A.~Olive and M.~Srednicki,
Phys.\ Lett.\ B {\bf 510} (2001) 236
[arXiv:hep-ph/0102098].


\bibitem{eoss}
J.~R.~Ellis, K.~A.~Olive, Y.~Santoso and V.~C.~Spanos,
Phys.\ Lett.\ B {\bf 565} (2003) 176
[arXiv:hep-ph/0303043].



\bibitem{cmssmwmap}
H.~Baer and C.~Balazs,
arXiv:hep-ph/0303114;
A.~B.~Lahanas and D.~V.~Nanopoulos,
arXiv:hep-ph/0303130;
U.~Chattopadhyay, A.~Corsetti and P.~Nath,
arXiv:hep-ph/0303201;
 C.~Munoz,
               hep-ph/0309346.


  \bibitem{nonu}
D.~Matalliotakis and H.~P.~Nilles,
  Nucl.\ Phys.\  B {\bf 435} (1995) 115
  [arXiv:hep-ph/9407251];  
  M.~Olechowski and S.~Pokorski,
  Phys.\ Lett.\ B {\bf 344}, 201 (1995)
  [arXiv:hep-ph/9407404];
V.~Berezinsky, A.~Bottino, J.~Ellis, N.~Fornengo, 
               G.~Mignola and S.~Scopel,
               {\em Astropart.\ Phys.}  {\bf 5} (1996) 1, 
               hep-ph/9508249;
               M.~Drees, M.~Nojiri, D.~Roy and Y.~Yamada,
               {\em Phys.\ Rev.} {\bf D 56} (1997) 276, 
               [Erratum-ibid.\ {\bf D 64} (1997) 039901], 
               hep-ph/9701219;
               M.~Drees, Y.~Kim, M.~Nojiri, D.~Toya, K.~Hasuko and 
               T.~Kobayashi,
               {\em Phys.\ Rev.} {\bf D 63} (2001) 035008, 
               hep-ph/0007202;
               P.~Nath and R.~Arnowitt,
               {\em Phys.\ Rev.} {\bf D 56} (1997) 2820, 
               hep-ph/9701301;
               J.~R.~Ellis, T.~Falk, G.~Ganis, K.~A.~Olive and M.~Schmitt,
  Phys.\ Rev.\ D {\bf 58} (1998) 095002
  [arXiv:hep-ph/9801445];
J.~R.~Ellis, T.~Falk, G.~Ganis and K.~A.~Olive,
  Phys.\ Rev.\ D {\bf 62} (2000) 075010
  [arXiv:hep-ph/0004169];
               A.~Bottino, F.~Donato, N.~Fornengo and S.~Scopel,
               {\em Phys.\ Rev.} {\bf D 63} (2001) 125003, 
               hep-ph/0010203;
               S.~Profumo,
               {\em Phys.\ Rev.} {\bf D 68} (2003) 015006, 
               hep-ph/0304071;
               D.~Cerdeno and C.~Munoz,
               {\em JHEP} {\bf 0410} (2004) 015, 
               hep-ph/0405057;
               H.~Baer, A.~Mustafayev, S.~Profumo, A.~Belyaev and X.~Tata,
               {\em JHEP} {\bf 0507} (2005) 065, 
               hep-ph/0504001.


\bibitem{nuhm}
J.~Ellis, K.~Olive and Y.~Santoso,
\PL B~{\bf 539}, 107 (2002)
[arXiv:hep-ph/0204192];
J.~R.~Ellis, T.~Falk, K.~A.~Olive and Y.~Santoso,
Nucl.\ Phys.\ B {\bf 652}, 259 (2003)
[arXiv:hep-ph/0210205].







\bibitem{eos1}
   J.~Ellis, K.~A.~Olive, and P.~Sandick,
   Phys.~Lett.~B {\bf 642} (2006) 389.

\bibitem{eos2}
   J.~Ellis, K.~A.~Olive, and P.~Sandick,
   JHEP {\bf 06} (2007) 079.
   
 \bibitem{mixed}
K.~Choi, A.~Falkowski, H.~P.~Nilles and M.~Olechowski,
  Nucl.\ Phys.\  B {\bf 718} (2005) 113
  [arXiv:hep-th/0503216]; 
    K.~Choi, K.~S.~Jeong and K.~i.~Okumura,
  JHEP {\bf 0509} (2005) 039
  [arXiv:hep-ph/0504037];
   M.~Endo, M.~Yamaguchi and K.~Yoshioka,
  Phys.\ Rev.\ D {\bf 72} (2005) 015004
  [arXiv:hep-ph/0504036];
   A.~Falkowski, O.~Lebedev and Y.~Mambrini,
     JHEP {\bf 0511} (2005) 034
     [arXiv:hep-ph/0507110];
   R.~Kitano and Y.~Nomura,
     Phys.\ Lett.\ B {\bf 631} (2005) 58
     [arXiv:hep-ph/0509039];
   R.~Kitano and Y.~Nomura,
     Phys.\ Rev.\ D {\bf 73} (2006) 095004
     [arXiv:hep-ph/0602096];
     A.~Pierce and J.~Thaler,
  JHEP {\bf 0609} (2006) 017
  [arXiv:hep-ph/0604192];
   K.~Kawagoe and M.~M.~Nojiri,
     [arXiv:hep-ph/0606104];
   H.~Baer, E.-K.~Park, X.~Tata and T.~T.~Wang,
     JHEP {\bf 0608} (2006) 041
     [arXiv:hep-ph/0604253];
      K.~Choi, K.~Y.~Lee, Y.~Shimizu, Y.~G.~Kim and K.~i.~Okumura,
  JCAP {\bf 0612} (2006) 017
  [arXiv:hep-ph/0609132];
  O.~Lebedev, V.~Lowen, Y.~Mambrini, H.~P.~Nilles and M.~Ratz,
  JHEP {\bf 0702} (2007) 063
  [arXiv:hep-ph/0612035].
  
   
   

\bibitem{WMAP} D.~N.~Spergel {\it et al.},
  [arXiv:astro-ph/0603449].


\bibitem{pbmz}
  D.~M.~Pierce, J.~A.~Bagger, K.~T.~Matchev and R.~j.~Zhang,
  Nucl.\ Phys.\  B {\bf 491}, 3 (1997)
  [arXiv:hep-ph/9606211].

\bibitem{tovey}
   D.~Tovey,
    Eur. Phys. J. direct C {\bf 4} CN4 (2002).


\bibitem{tovey2}
   D.~Tovey, private communication.

\bibitem{ILCexecsum}
  J.~Brau {\it et al.},
  {\it International Linear Collider reference design report. 1: Executive
  summary. 2: Physics at the ILC. 3: Accelerator. 4: Detectors}, SLAC-R-857 (2007).

 

\end{thebibliography}
\end{document}